\documentclass[preprint,authoryear,12pt]{elsarticle}
\usepackage{graphicx}
\usepackage{amssymb}
\usepackage{amsthm}
\usepackage{amsmath}
\usepackage{xcolor}

\usepackage[T1,T2A]{fontenc}
\usepackage[utf8]{inputenc}

\usepackage[english]{babel}
\usepackage{nicefrac}
\usepackage[normalem]{ulem}
\usepackage{makecell}
\usepackage{graphbox}

\usepackage{float}
\floatstyle{plaintop}
\restylefloat{table}
\usepackage{hyperref}

\makeatletter
\DeclareRobustCommand\bigop[1]{%
  \mathop{\vphantom{\min}\mathpalette\bigop@{#1}}\slimits@
}
\newcommand{\bigop@}[2]{%
  \vcenter{%
    \sbox\z@{$#1\sum$}%
    \hbox{\resizebox{\ifx#1\displaystyle1.1\fi\dimexpr\ht\z@+\dp\z@}{!}{$\m@th#2$}}%
  }%
}
\makeatother

\newcommand{\minmax}{\DOTSB\bigop{\mathrm{m_{in}^{ax}}}}

\journal{Ecological Modelling}

\begin{document}
\begin{frontmatter}

\title{``Perchance to dream?'': Assessing effect of dispersal strategies on the fitness of expanding populations}

\author{N.I. Markov$^1$, E.E. Ivanko$^2$}


\address{$^1$Institute of Plant and Animal Ecology\\
$^2$Institute of Mathematics and Mechanics\\ 
Ural Branch of the Russian Academy of Sciences\\ Ekaterinburg, Russia
}

\begin{abstract}

Unraveling patterns of animals' movements is important for understanding the fundamental basics of biogeography, tracking range shifts resulting from climate change, predicting and preventing biological invansions. Many researchers have modeled animals' dispersal studying their behavior under the assumptions of some movement strategies pre-determined or affected by some external factor(s) but none of them have compared the efficiency of different dispersal strategies in providing population survival and fitness. We hypothesize that 1) successful expansion could result from some evolutionary stable strategy (ESS) and 2) such strategy could be based particularly on deferred gain, when animals invest in travel to reach some high-quality habitat (``habitat of dream''). Using simulation model we compare the ecological success of three strategies: i) ``Smart'' -- choosing the locally optimal cell; ii) ``Random'' -- random movement between cells without taking into account the quality of the environment; iii) ``Dreamer'' -- movements that aims to find ``a habitat of dream'' with  quality much higher than that of the initial and neighboring cells. The population fitness was measured as survival rate, dispersal distance, accumulated energy and quality of settled habitat. The most general conclusion is that while survival and wealth of the population is affected presumably by overall habitat quality, the dispersal depends mainly on the behavioral strategy. The ``Dreamer'' strategy or the strategy of deferred gain belongs to the Pareto frontier in the {\it Fitness$\times$Dispersal} space but only in optimal and suboptimal habitat and in the relatively mild climate. 
\end{abstract}

\begin{keyword}
Dispersal \sep habitat quality \sep search strategy \sep Levy walk \sep survival \sep population fitness

\end{keyword}

\end{frontmatter}


\section{Introduction}
\label{S:1}

Animals differ from plants by their ability to move actively and choose an optimal environment. Animals move for many reasons: to acquire resources; to avoid predators and other agents of mortality; to avoid competition (e.g. natal dispersal); and to be near conspecifics for mating and other social interactions (conspecific attraction). Clearly, the various functions of movement are related to survival and reproduction, and the parameters that govern movement are therefore subject to natural selection.

One of the most interesting types of animals' movements are the so-called ``long-distance movements'' of individuals from the places of their birth or release. During such movements an organism can disperse far beyond the region of its origin through a variety of landscapes and then either return back or settle in a very new environment. Long-distance movements or long-distance dispersal (LDD) has gained most attention in the studies of passive dispersal of plants (\cite{Nathan2008movement}). A number of reports has been presented for invertebrates and birds (\cite{Earl2016Characteristics}). Recent advances in satellite-tracking techniques increased the number of such reports for mammals (\cite{Fuglei2019Arctic,Hindell2000Long, Weller2016First}). Studies of ungulates' behavior have shown that despite rare long-distance movements have been observed for many species. For example, for wild boar ({it Sus scrofa}) most of animals dispersed not more than 20 km from their place of birth, but some individuals moved more than 60 or even more than 80 km (\cite{Truve2004Dispersal}).

Approaches to study LDD were classified in \cite{Nathan2003Methods} and mathematical modeling was listed as one of the possible ways to understand this phenomenon together with a set of observational methods. Generally, model-based studies of animals' movements address the following main problems:

1. Approximation of the empirical movement (presumably telemetry) data with mathematically defined processes, like random (correlated or uncorrelated, biased or unbiased) walk or Levy walk (e.g. reviews \cite{Hawkes2009Linking,Lewis2015Mathematics});

2. Using mathematical approaches (like, for example, partial differential equations) to model population dynamics accounting for dispersal and range shifts, particularly during biological invasions (reviewed in \cite{Lewis2015Mathematics});

3. Effect of habitat heterogeneity and population dynamics on the process of dispersal and patterns of animals' movements (\cite{Hawkes2009Linking, Barton2009evolution, Hiebeler2004Competition});

4. Habitat selection during dispersal and the factors affecting it (\cite{Davis2004effect, Stamps2005Search, Stamps2007Someplace}).

Thus the models of animals' movements study the behavior under the assumptions of some movement strategy, but none of them (to our knowledge) address the problem of comparing the efficiency of different strategies. \cite{Lewis2015Mathematics} wrote ``long-distance dispersal can be the driving force for population spread if populations introduced at low densities thrive and do not go extinct''. But what is the chance for a population not to extinct if it implements LDD? -- that is the question! On one hand the wide distribution of this phenomenon among living organisms (\cite{Nathan2005Long-distance, Sutherland2000Scaling, Teitelbaum2015How}) indicates that this could be (under certain conditions) an important part of species' life history. On the other hand the relative rarity of such events (in compare with short distance dispersal) (\cite{Wilson2009Something}) shows that animals moving long distances most probably are subject to higher mortality and thus the selection against long-distance dispersal could take place.

In this study we hypothesize that 1) dispersal success could result from some evolutionary stable strategy (ESS) of maximizing the population fitness and 2) that this strategy could be based on deferred gain, when animals invest in travel to reach some high-quality habitat (``habitat of dream''). The last hypothesis is rooted in the concept of natal habitat preference induction (\cite{Davis2004effect}) which states that the experience with the natal habitat shapes the habitat preferences of the individuals that disperse particularly after releases in new areas (\cite{Stamps2007Someplace}).

Checking these hypotheses using observational methods is practically impossible since we always have to guess what animals have in their minds while choosing one or another way of expansion: e.g., is it a random choice or it is pre-determined by some ESS. We address this problem via programming individuals search strategies. In other words, we tell the modeled animals (animats) (\cite{Wilson1991animat}) how to behave and then check which behavior would result in higher fitness of the whole population.  

We compare the ecological (short-term) success of three strategies:

-- ``Smart'' -- choosing the locally optimal cell (a cell with the highest quality among the neighboring units);

-- ``Random''  -- random movement between cells without taking into account the quality of the environment;

-- ``Dreamer'' -- movements aims to find ``a cell of dream'' with  quality much higher than that of the initial and neighboring cells.

Comparing the strategies, we address three main questions:

1) What is the relative success of each strategy in bad, medium and good habitat?

2) How the success of each strategy could be affected by the harshness of environment (particularly, environmental seasonality)?  

3) Could the ``Dreamer'' strategy provide sufficient fitness for the population implementing this strategy?

\section{Model}

We modeled the movements of organisms taken  from optimal habitat and released  in a new environment where the spatial distribution of habitat patches and quality of habitat is not known by them. There is a number of examples of such (intentional or non-intentional) translocations of animals by humans, which resulted in extinction or successful expansion of newly established populations (\cite{Bertolero2007Assessing,Bremner-Harrison2004Behavioural,Lodge2003Biological}).

\subsection{Map}

\begin{figure}
    \centering
    \includegraphics[width=0.35\textwidth]{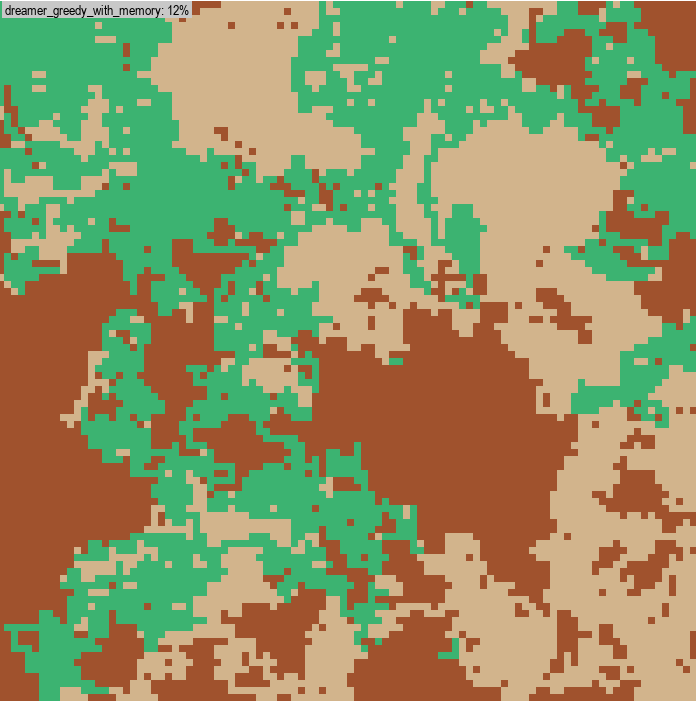}\\
    \vspace{3mm}
    \includegraphics[width=0.5\textwidth]{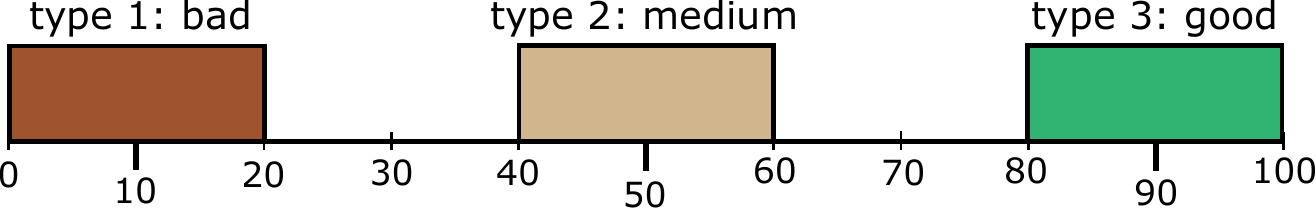}
    \caption{An example of the $100\times100$ experiment maps  generated by Algorithm 1. All three strata occupy equal area: $S_{bad}/S_{norm}/S_{good}=1/1/1$ (dark-brown is for ``bad'', light-brown -- for ``medium'' and green -- for ``good''), ``patchness'' is equal to 5. The scale shows the connection between cell types (strata) and energy (attractiveness)}
    \label{fig:map_example}
\end{figure}

The map in the experiments is a square divided into $100\times100$ square cells. Each cell belongs to one of three strata: ``bad'', ``medium'' or ``good''. For the details on the process of map construction see Algorithm 1 in \ref{alg_sec_map} and a map example in Fig.\ref{fig:map_example}. Note that Algorithm 1 allows to produce maps with any given area ratio $S_{bad}/S_{medium}/S_{good}$ and different ``pathchness''. In our experiments we used three types of map: ``bad'' -- with $S_{bad}/S_{medium}/S_{good} = 0.67/0.3/0.03$, ``medium'' -- with $S_{bad}/S_{medium}/S_{good} = 0.22/0.56/0.22$ and ``good'' -- with $S_{bad}/S_{medium}/S_{good} = 0.03/0.3/0.67$.

\subsection{Energy}

There are two types of energy in the model: energy of the animats (which positive level is necessary for each animat to live and to move) and energy (or attractiveness) of the cells on the map (which serves as a source of energy for the animats). The latter one plays the role of food for the animats; it does not deplete but can change over time depending on season.

\subsubsection{Energy of cells}

Each cell is assigned a random energy value (attractiveness) according to its strata: from 1 to 20 for ``bad'', from 40 to 60 for ``medium'', and from 80 to 100 for ``good''. 

\begin{figure}
    \centering
    \includegraphics[width=0.35\textwidth]{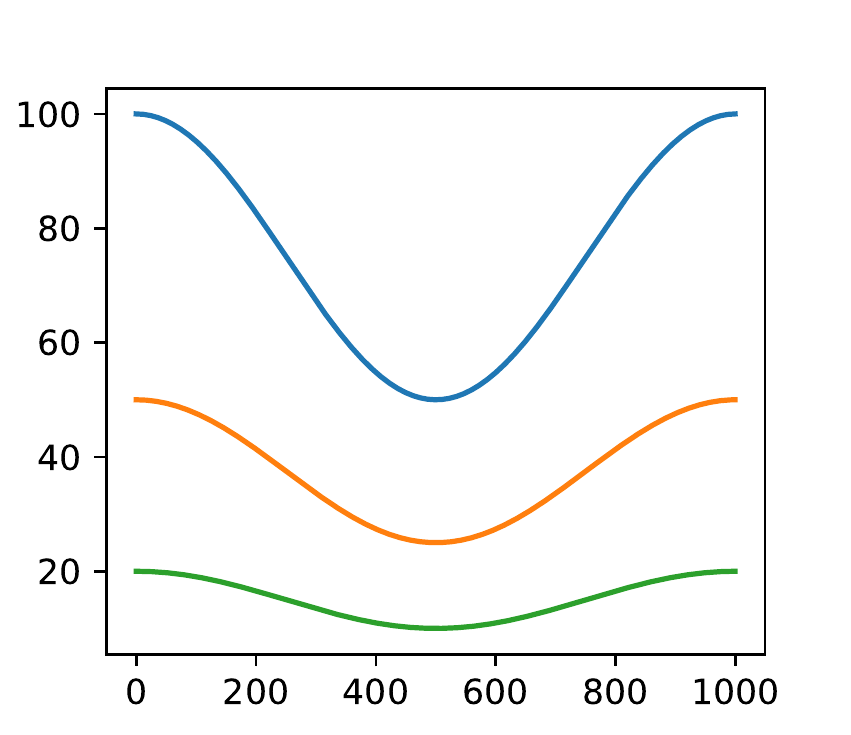}
    \caption{An example of ``seasonal'' variation of the cells' energy (attractiveness) during a 1000 steps experiment cycle. The variation is presented for 3 cells with initial energies 100, 50, 20 and ``winter harshness'' equal to 2}
    \label{fig:energy_sine_example}
\end{figure}

The energy of the cells follow a ``weather conditions'' sine wave during a 1000 steps cycle: first getting worse to model ``autumn'' and ``winter'' and then recover to ``spring'' and ``summer''. At the worst time (in the middle of winter) the attractiveness of each cell drops to the lowest value $q/w$, where $q$ is the initial attractiveness of this cell and $w$ is the experiment parameter -- ``winter harshness''. The formal expression for the attractiveness (energy) of a cell with the initial energy $q$ at the step $j$ (varying from 1 to 1000) is 
\begin{equation}
\label{eq:weather_sine}
    \frac{q-\tfrac{q}{w}}{2}\cdot cos\left(\frac{2\pi j}{1000}\right)+\frac{q+\tfrac{q}{w}}{2},
\end{equation}
(see examples of these waves for different $q$ in Fig. \ref{fig:energy_sine_example}).

\subsubsection{Energy of animats}

Each animat constantly acquires energy from the cells it visits  and spends it for metabolism and movements. In our experiments, we assume that an animat spends $E^m=25$ units of energy for each shift to a neighbor cell
(half as much as the average initial energy of the cells in the ``medium'' strata). If an animat cannot make a move at the current step, it spends $E^m/2=12.5$ units of energy for just keeping homeostasis (twice as low as the energy for a move). 
Each newly created animat is provided with $E^0=10E^m$ units of energy (so in can make 10 moves without any energy input). After arriving at a new cell an animat receives the amount of energy given by \eqref{eq:weather_sine} for this cell at the current step of the ``weather'' sine wave. If the energy balance of an animat drops below zero, the animat dies.

\subsection{Movement}

We test 3 search strategies using the movement tactics based on Levy-walk. In our variant of the Levy-walk model, the animats choose
each next direction randomly but restricted to the forward semiplane with respect to the vector of the previous run (the first direction is selected without any restrictions). This peculiarity brings an element of the correlated random walk to our model. The number of successive steps in each direction is chosen in accordance with the Levy-walk model with $\mu=2$ (\cite{Viswanathan1999Optimizing}), but does not exceed 50 (the expectation of the animat's position is the map's center, which lies 50 or little more cells away from the edges in any direction). Reaching an edge of the map naturally leads to a re-initialization of the Levy-walk without the ``semiplane restriction''.

The movements include a concept of memory: the animats do not shift to the cells they visited during the previous $r$ steps. For example, if $r=3$ and an animat starts from the cell $A$, shifts to $B$ then to $C$ and remains there, then the memory queue goes through the following states: $AAA$ -- $AAB$ -- $ABC$ -- $BCC$ -- $CCC$; after the first step the animat is not allowed to return to $A$, after the second -- to both $A$ and $B$, after the third -- to $A$, $B$, and $C$, after the fourth -- only to $B$ and $C$. In our experiments, the memory size was chosen equal to 10: little more than the number of neighbors of each inner cell in a square grid.

It is important to note that in our model a moving animat can recognize strata (good, medium or bad) without entering it (so the animat may refuse to enter a cell of a worse type), but the animats cannot distinguish ``by eye'' the energy attractiveness of two cells belonging to one strata.  

The three search strategies studied in the experiments were:

1. ``Random'' or ``no determination by environment''. 
In this strategy each animat randomly chooses a direction and a length from a Levy-walk distribution and follows the chosen direction until the chosen length is achieved or until it comes to an edge of the map. The ``Random'' animats do not take into account the strata or energy of the cells they pass through: for example, an animat can move from a cell with quality 5 to a cell with quality 1 (for the details on the movement of ``Randoms'' see Algorithm 2 in \ref{sec:b_algorithms}). We hypothesize that in a completely new environment such behavior could occur due to partial inability of organisms to analyze efficiently the whole volume of incoming information.

2. ``Smart'' or ``full determination''. While following the same Levy-walk movement pattern a ``Smart'' animat shifts to the cell proposed by its current movement tactics only if the type of this new cell is not worse than the type of the current cell (see Algorithm 3 in \ref{sec:b_algorithms} for the details). For example, a ``Smart'' animat cannot move from a ``good'' habitat to a ``medium'' one, but it can move from a cell with energy 55 to a cell with energy 45 (within the category ``medium''). 
If the next step in the Levy-walk movement model leads to the decrease of the habitat quality, the successive Levy walk is stopped and reinitialized (with the choice of a new direction and ``flight'' length). From mathematical point of view this strategy is a {\it greedy local search} heuristic (\cite{metaheuristics2010handbook}) which is a well-known and extensively used instrument in the field of optimization.

3. “Dreamer” or “partial determination”. Each ``Dreamer'' animat has an idea of ``dream habitat'': a cell with enough attractiveness to stop the search and to rest contently.
Together with remembering the attractiveness of its dream, a ``Dreamer'' estimates the expectation to achieve it: this expectation starts with 100\% and reduces with each step where the dream is not found (see the details in \ref{sec:b_algorithms} after Algorithm 3). A ``Dreamer'' stops when the attractiveness of the ``dream'' multiplied by the expectation to achieve it becomes less than the attractiveness (energy) of the current cell (given that the probability to achieve the current cell is evidently 1) (see Algorithm 4 in \ref{sec:b_algorithms} for the formal expression of the stop condition). 
Note that due to the changing ``weather conditions'' a ``dreamer''  that has already taken the decision to rest content with the current cell can continue moving since the current cell attractiveness (energy) becomes too small to trump its dream.
This strategy as well has a clear analogue in the theory of optimization – {\it simulated annealing} technique (\cite{metaheuristics2010handbook}), where the intention of an agent to explore the space of the problem states gradually decreases in time until the agent finally satisfies with its current achievements.

\subsection{Metrics}
\label{sec:metrics_general}
The following parameters were assessed in each experiment:

-- proportion of the survived individuals (S);

-- dispersal from the starting points (D) (Euclidean distance between the initial and final cells);

-- energy accumulated by the animats during the experiment (E);

-- quality of the cells occupied at the end of the experiment (Q);

-- the product of the accumulated energy and cell quality further denoted as {\it Fitness} showing the ability of an individual not just to survive but to occupy good habitat and to accumulate energy for the future (EQ).

These metrics aimed at the characterization of homogeneous 100-animat groups or ``populations'' (where each member follows the same search strategy) at the end of the 1000-step experiment cycles. Each metric was computed in two variants: average over the individuals that managed to survive until the end of experiment and sum over all the members of the population (including those who died). While the former variant describes the survivors only, the latter one characterizes the populations as wholes (note that it is proportional to the average over {\it all} the members of the population). All the metrics were additionally normalized into $[0,1]$ to compare their relative manifestation within different strategies. The rigorous expressions for the metrics are given in \ref{sec:b_metrics}, see equations \eqref{b_aver_n}--\eqref{eq:normalized2} for the details).
 







\subsection{Experiment}

The 1000 step experiment cycle was repeated 30 times for each of the three types of 100-animat groups (``Randoms'', ``Smarts'' and ``Dreamers''), each of the three types of map (``bad'', ``medium'', ``good''), and each of the five types of winter harshness ($w=2,4,6,8,10$). Thus, the final data set consisted of 30 independent estimates of the metrics (S,D,E,Q,EQ) for each combination of terms (search strategy, map and winter harshness) (see  \ref{sec:b_metrics} for the details). 
A new map was generated every time after ``Randoms'', ``Smarts'' and ``Dreamers'' had used it each once in the same experiment conditions. Constant and variable parameters of the experiments are summarized in Table \ref{tab:all_parameters}.
The computational experiments were conducted using Python3, the code is available at GitHub (\cite{software}).

\begingroup
\setlength{\tabcolsep}{10pt} 
\renewcommand{\arraystretch}{1.3} 
\begin{table}
    \centering
    \vspace{5mm}
    \small
    \begin{tabular}{c|c}
       {\bf Terms in the model}  & {\bf Value} \\
        \hline\multicolumn{2}{c}{Constants}\\\hline
        Map size & $100\times 100$\\
        Energy of cells in good habitat & 80-100\\
        Energy  of  cells in medium habitat	& 40-60\\
        Energy  of  cells in bad habitat & 0-20\\
        Energy of ``dream habitat'' & 10000\\
        Initial energy of animat & 250\\
        Cost of animat movement & 25\\
        Cost of animat staying	& 12.5\\
        Duration of cycle & 1000 steps\\
        Animat memory size	& 10 steps\\
        \hline\multicolumn{2}{c}{Variables}\\\hline
        Proportions of habitat in: & (bad/medium/good) \\
        -- ``bad'' map & 0.67/0.3/0.03\\
        -- ``medium'' map & 0.22/0.56/0.22\\
        -- ``good'' map & 0.03/0.3/0.67\\
        Winter harshness & 2, 4, 6, 8, 10\\
        Search strategy & ``Random'', ``Smart'', ``Dreamer''
    \end{tabular}
    \caption{Constant and variable terms of the model}
    \label{tab:all_parameters}
\end{table}
\endgroup

\section{Results}


\subsection{``Bad'' map (model of bad habitat  prevalence)}

(Fig.\ref{fig:2d_aver}a,d,g,j and Fig.\ref{fig:2d_sum}a,d,g,j)

On the ``bad'' maps, ``Smarts'' considered as a whole population (i.e. using the summary metrics),  perform much better than the other strategies. All their metrics are the highest and almost independent of the winter harshness except for E, which decreases in a non-linear manner with the increase of the winter harshness. The ``Smart'' strategy animats tend to find locally optimal habitat patches quickly and are able to withstand the ``seasonal'' changes in cells quality by moving (or staying) within the discovered favorable habitat patches (thus maximizing the energy accumulation and survival). The comparison of “Smarts” with the other strategies basing on the average metrics on the ``bad'' maps makes sense only in the case of mild winters: in harsh winters the low survival rate of “Dreamers” and “Randoms” increases the bias of the corresponding average values.
 Though in mild winters the average metrics of the ``Smarts'' generally repeat the corresponding summary metrics, there are two exceptions: 1) the average dispersal of both ``Dreamer'' and ``Random'' survivors is considerably larger than the one of ``Smarts''; 2) ``Dreamers'' and ``Smarts'' settle the habitat of almost the same quality.

``Randoms'' are the worst in the bad habitat -- they demonstrate the lowest values of all metrics (both summary and averaged, except for the dispersal in the mildest winter $w=2$). Their survival does not exceed 10\% and the average quality of the cells occupied by a few survivors at the end of the experiments does not exceed 20. The strategy of random movements does not allow the animats to find patches of good or medium habitat in the ``bad'' map and they die quickly due to high cost/gain energy ratios. The increase of winter harshness from 2 to 4 results in almost zero survival rate thus the further increase of $w$ does not significantly affect the metrics.

``Dreamers'' demonstrate an intermediate performance but in the summary metrics are closer to ``Randoms'' than to ``Smarts''.  “Dreamers” metrics decrease with the increase of winter harshness, possibly due to the strong decrease of survival in harsh winters. According to the values of average cell quality the few survivors are those who managed to reach good and medium habitat.

\subsection{``Medium'' map (model of medium habitat prevalence)} 

(Fig.\ref{fig:2d_aver}b,e,h,k and Fig.\ref{fig:2d_sum}b,e,h,k)

In medium habitat  ``Smarts'' show almost no mortality (S above 0.9), E and Q values (both average and sums) are the highest, and D is the lowest observed among the three strategies. The effect of winter harshness is insignificant for all metrics except for E, which is negatively correlated with the winter conditions. 

``Randoms'' perform almost as good as the other strategies in mild winters in terms of survival. Also in mild winters their dispersal is the highest. Their E and Q metrics are however the lowest ones. All metrics except dispersal decrease linearly with the increase of winter harshness. Thus the strategy of random movement allows the animats to survive well in the medium habitats which quality does not change strongly within the season. Even in very harsh winters their survival does not fall below 40\% which shows that this strategy could theoretically help animals to survive and to establish viable populations in suboptimal environment, however the probability of it stays much lower than for ``Dreamers'' and ``Smarts''.

In differ with the ``bad'' maps, in the ``medium'' maps ``Dreamers'' perform much better than ``Randoms'' in terms of S, E and Q. Their values of the summary population D are higher than for the other strategies (except for ``Randoms'' at $w$=2) and this index does not change significantly with winter harshness, while the average D of the survivors almost reaches the ``Randoms'' level. The effect of winter on the ``Dreamers'' S and Q metrics is not very high (about 20\%) but noticeable. The E index decreases twice with the 5-fold increase in winter harshness. Thus in medium habitats the implementation of the ``Dreamer'' strategy results in relatively high population success in terms of survival, expansion and occupation of good habitat.

\subsection{``Good'' map (model of good habitat prevalence)} 

(Fig.\ref{fig:2d_aver}c,f,i,l and Fig.\ref{fig:2d_sum}c,f,i,l)

In good habitat the survival of ``Smarts'', ``Dreamers'' and ``Randoms'' is above 98\% showing almost no mortality. In this case the survival rate does not affect the average metrics and they are equal to the corresponding summary ones.

``Randoms'' are the most successful in dispersal (which is not a surprise given their dispersal is not intermitted by any decisions), but least successful in energy accumulation and the quality of habitat occupied by the animats at the end of the experiments.

``Dreamers'' perform better than the other strategies in terms of Q, they are as good as ``Smarts'' in energy accumulation (E). Their dispersal D strongly increases with winter harshness approaching the values typical for ``Randoms''. The dispersal of ``Randoms'' and ``Smarts'' is in fact not affected by the winter harshness. For all the strategies, the E index decreases with winter harshness, while the slight decrease in Q was observed for ``Randoms'' and ``Dreamers'' but not for ``Smarts''. Thus in favorable environment all the strategies are almost equally successful. 

\begingroup
\setlength{\tabcolsep}{5pt}

\begin{figure}[H]
    \centering
\begin{tabular}{cccc}
& {\footnotesize``bad'' map} & {\footnotesize``medium'' map} & {\footnotesize``good'' map} \\
\rotatebox[origin=c]{90}{\footnotesize \qquad survival ($\mathbf{S}$)\qquad } &
\makecell{(a)\\\includegraphics[align=c,width=0.3\textwidth]{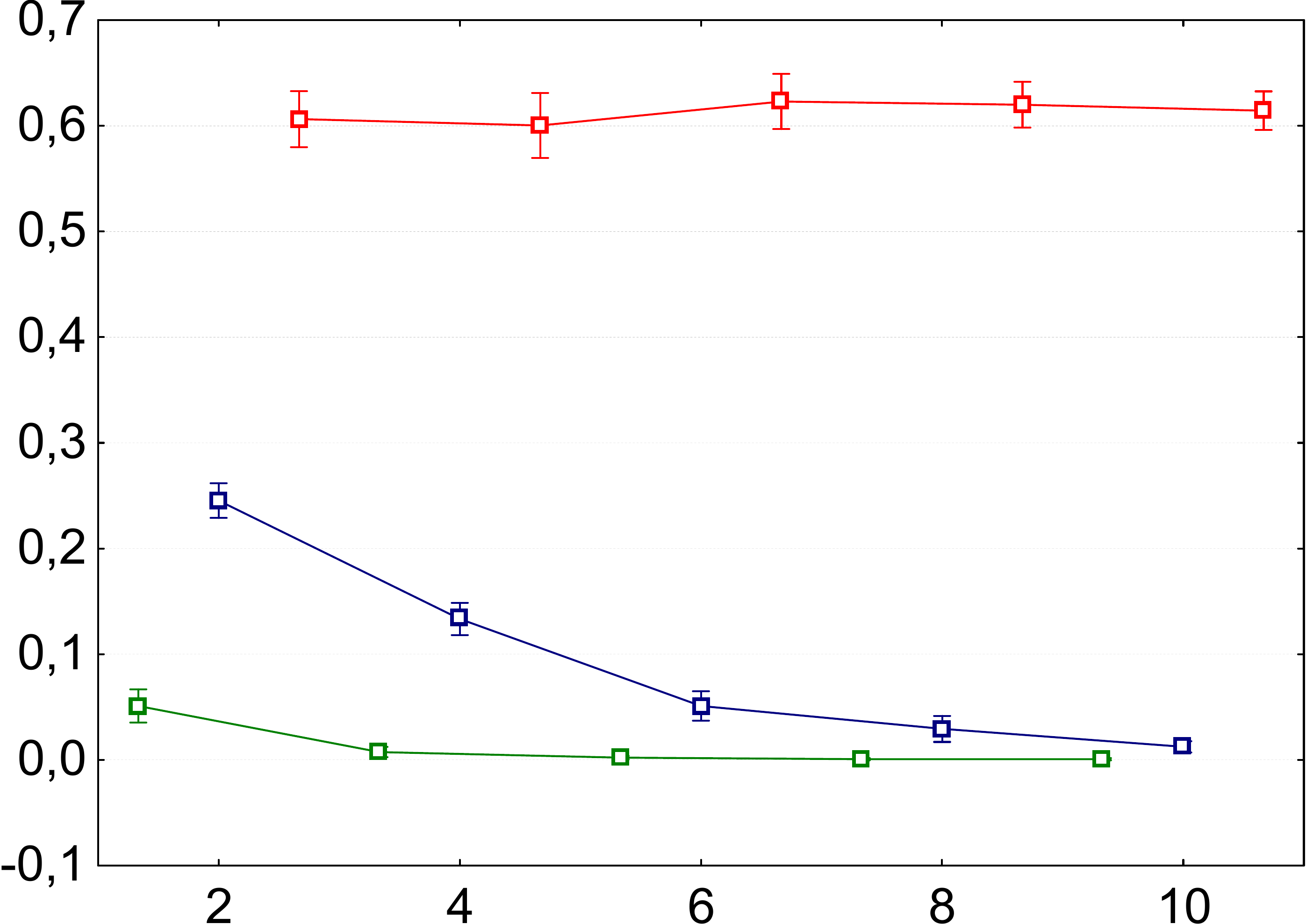}}
     &
\makecell{(b)\\\includegraphics[align=c,width=0.3\textwidth]{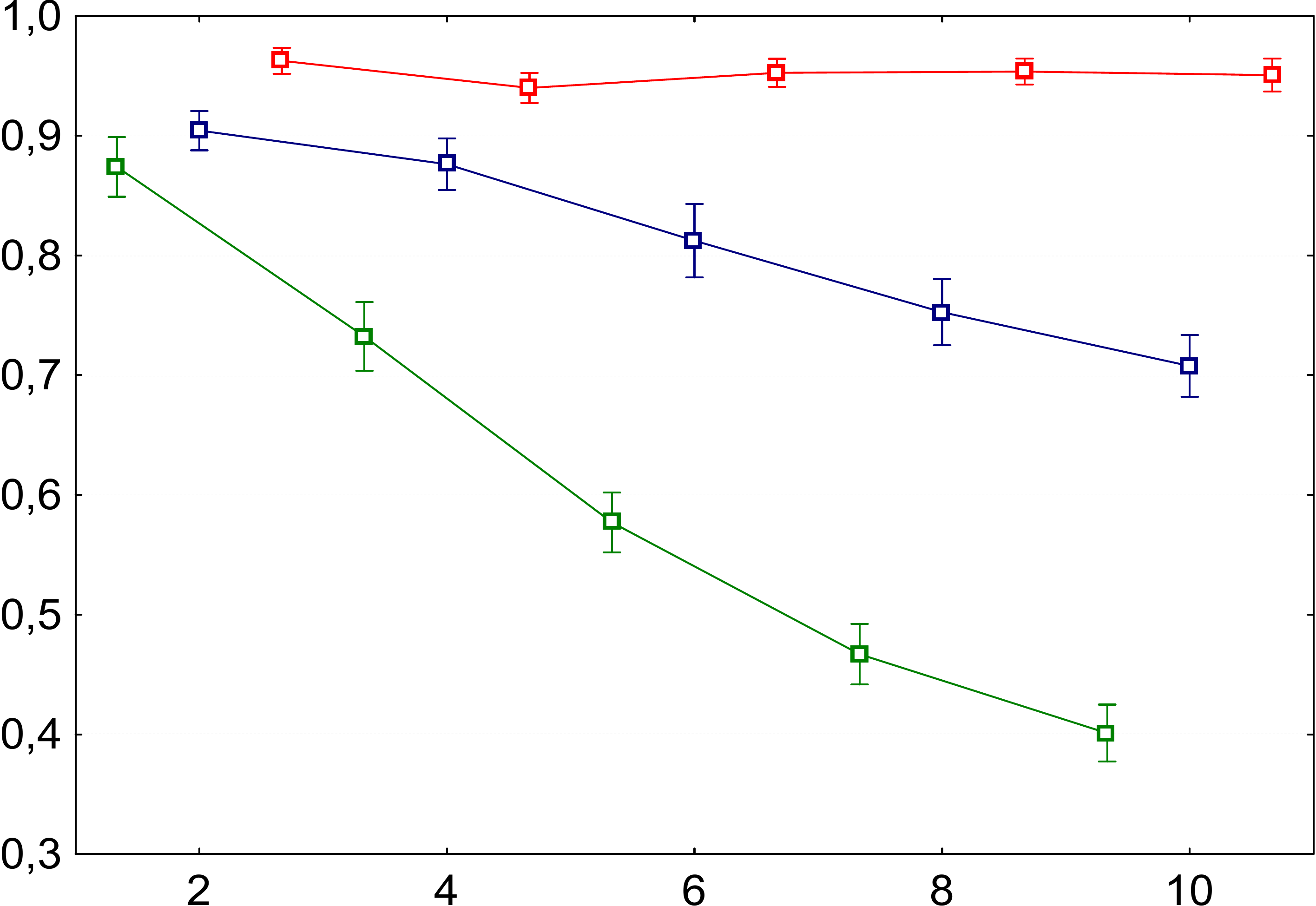}}
     &
\makecell{(c)\\\includegraphics[align=c,width=0.31\textwidth]{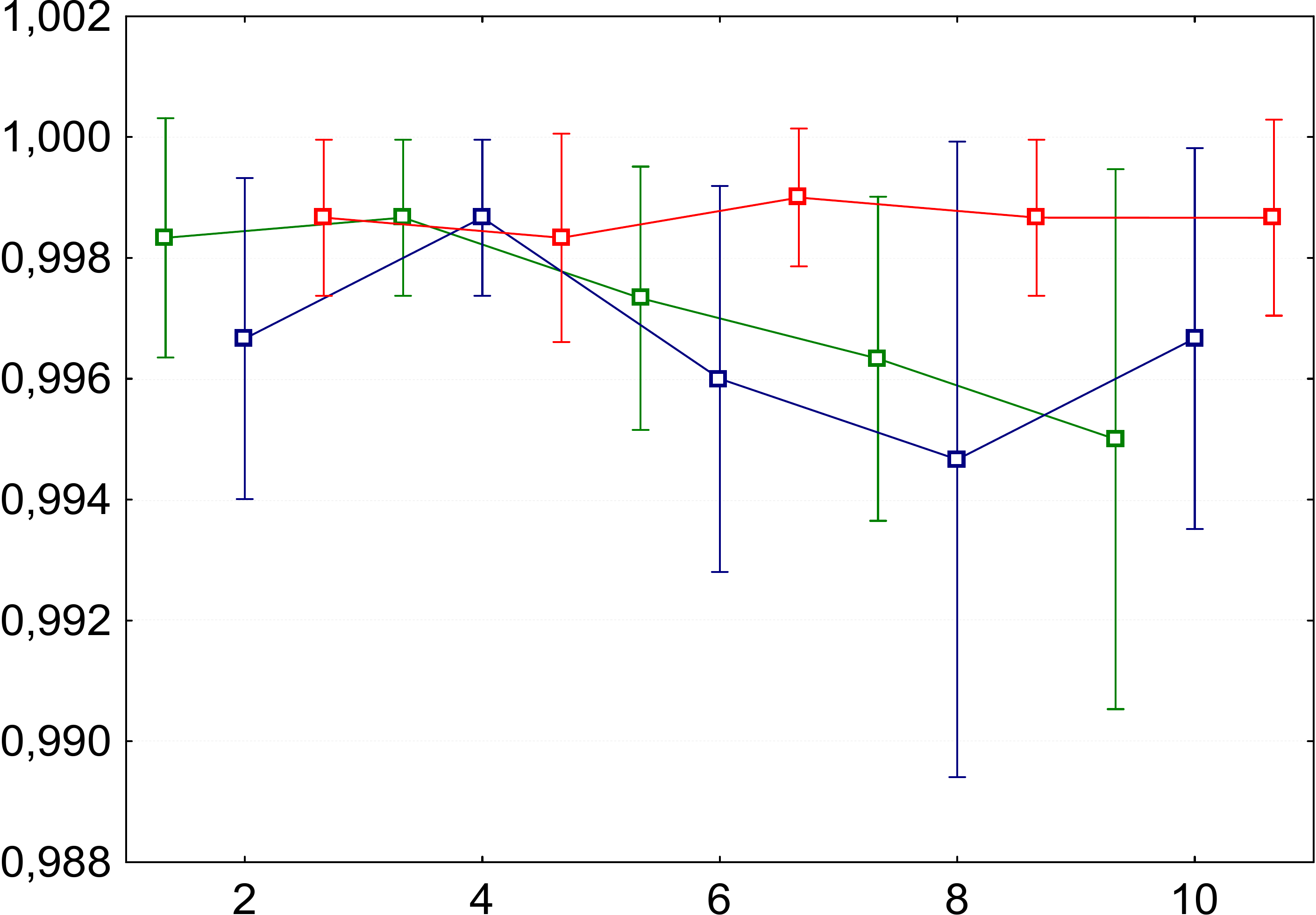}}\\
\rotatebox[origin=c]{90}{\footnotesize \qquad dispersal ($\mathbf{D}_A$)\qquad } &
\makecell{(d)\\\includegraphics[align=c,width=0.3\textwidth]{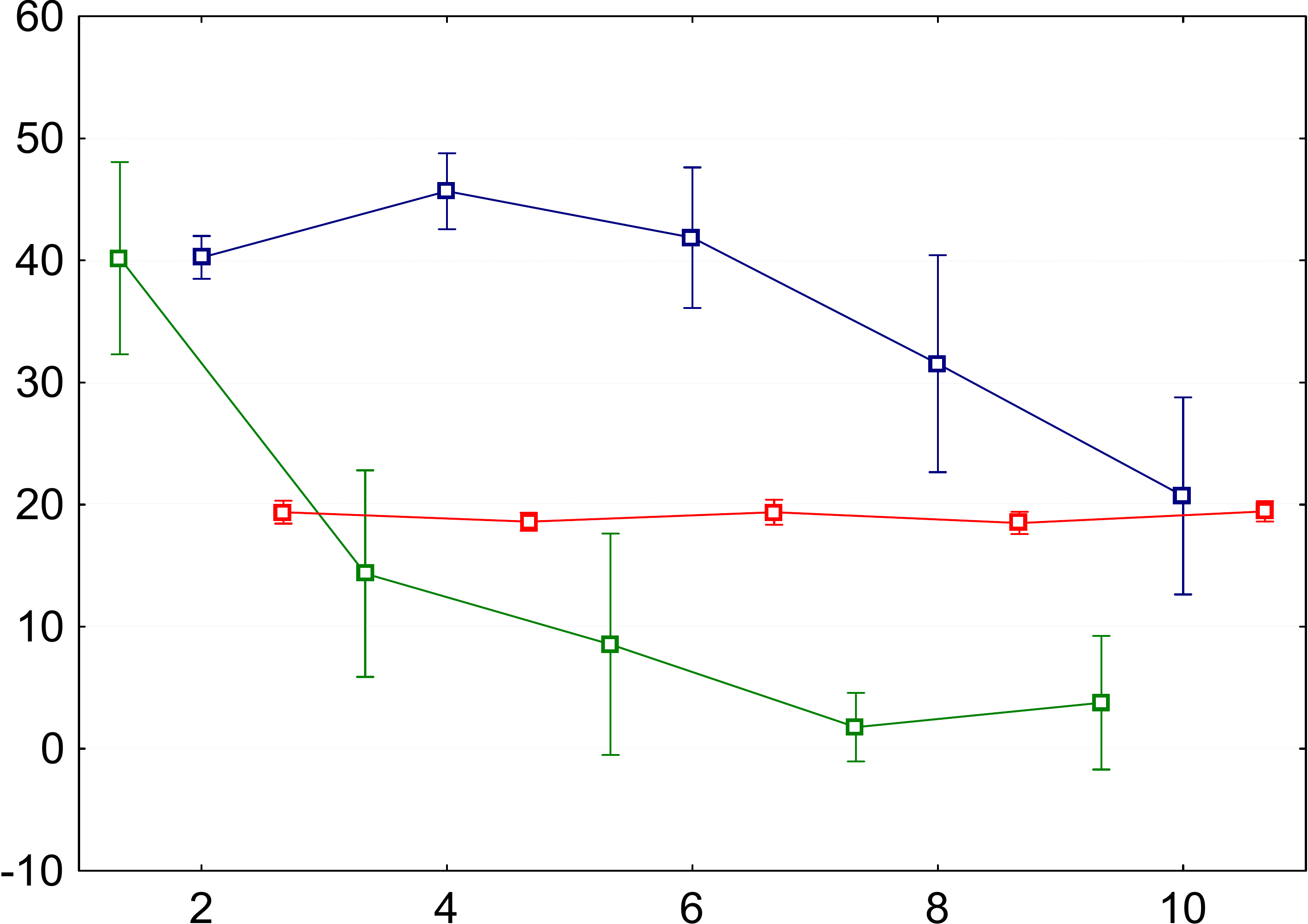}}
     &
\makecell{(e)\\\includegraphics[align=c,width=0.3\textwidth]{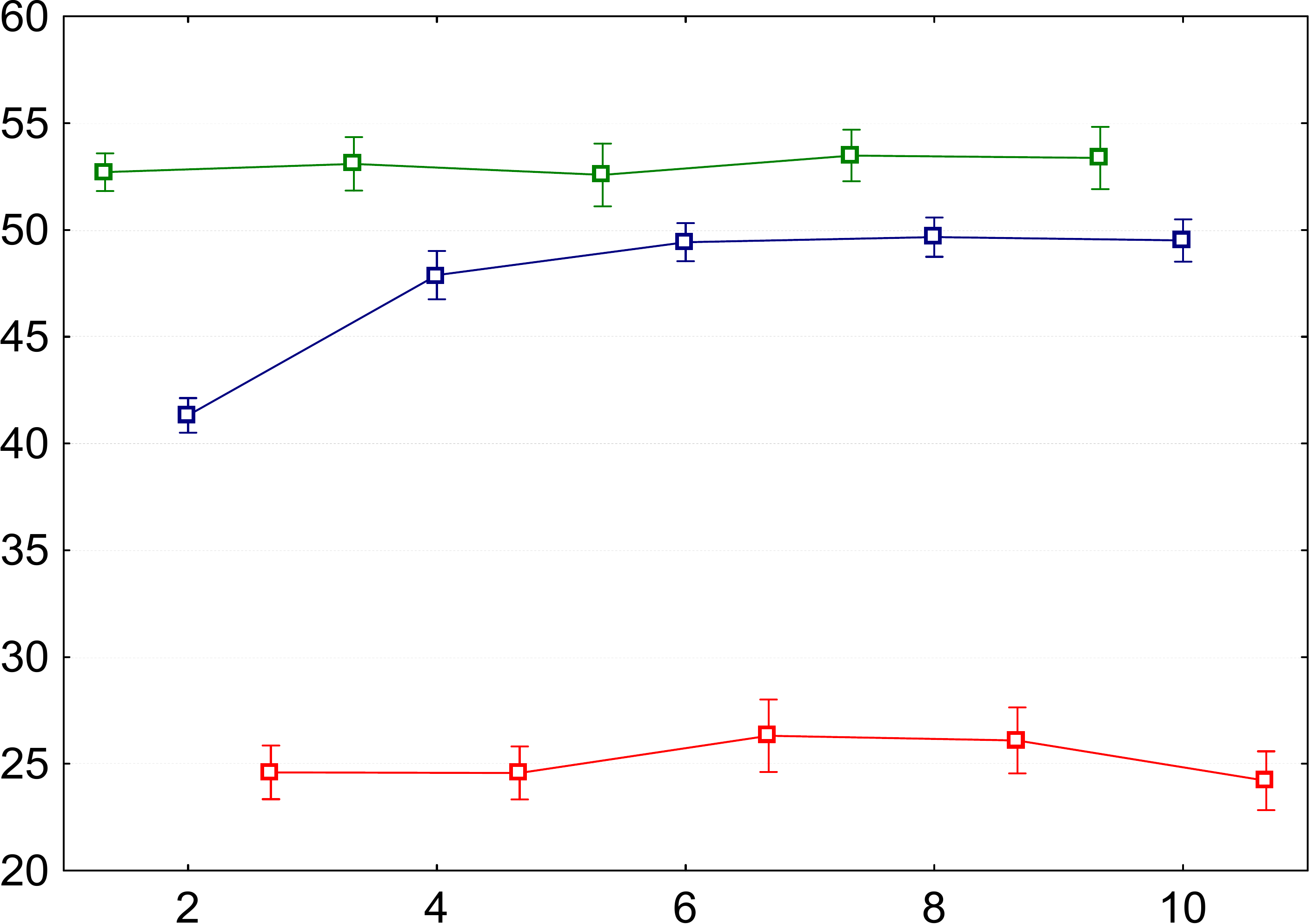}}
     &
\makecell{(f)\\\includegraphics[align=c,width=0.3\textwidth]{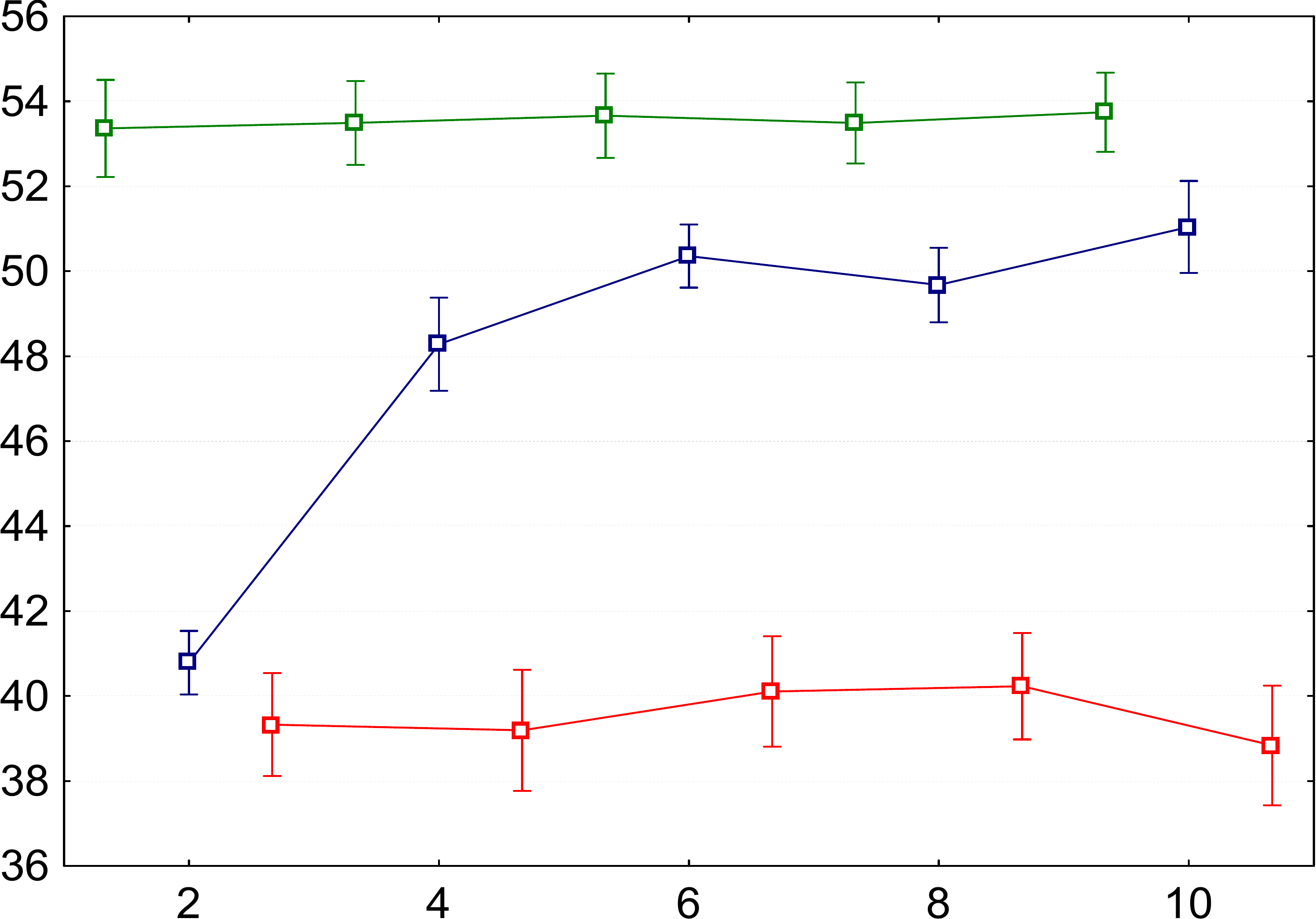}}\\
\rotatebox[origin=c]{90}{\footnotesize \qquad  energy ($\mathbf{E}_A$)\qquad } &
\makecell{(g)\\\includegraphics[align=c,width=0.31\textwidth]{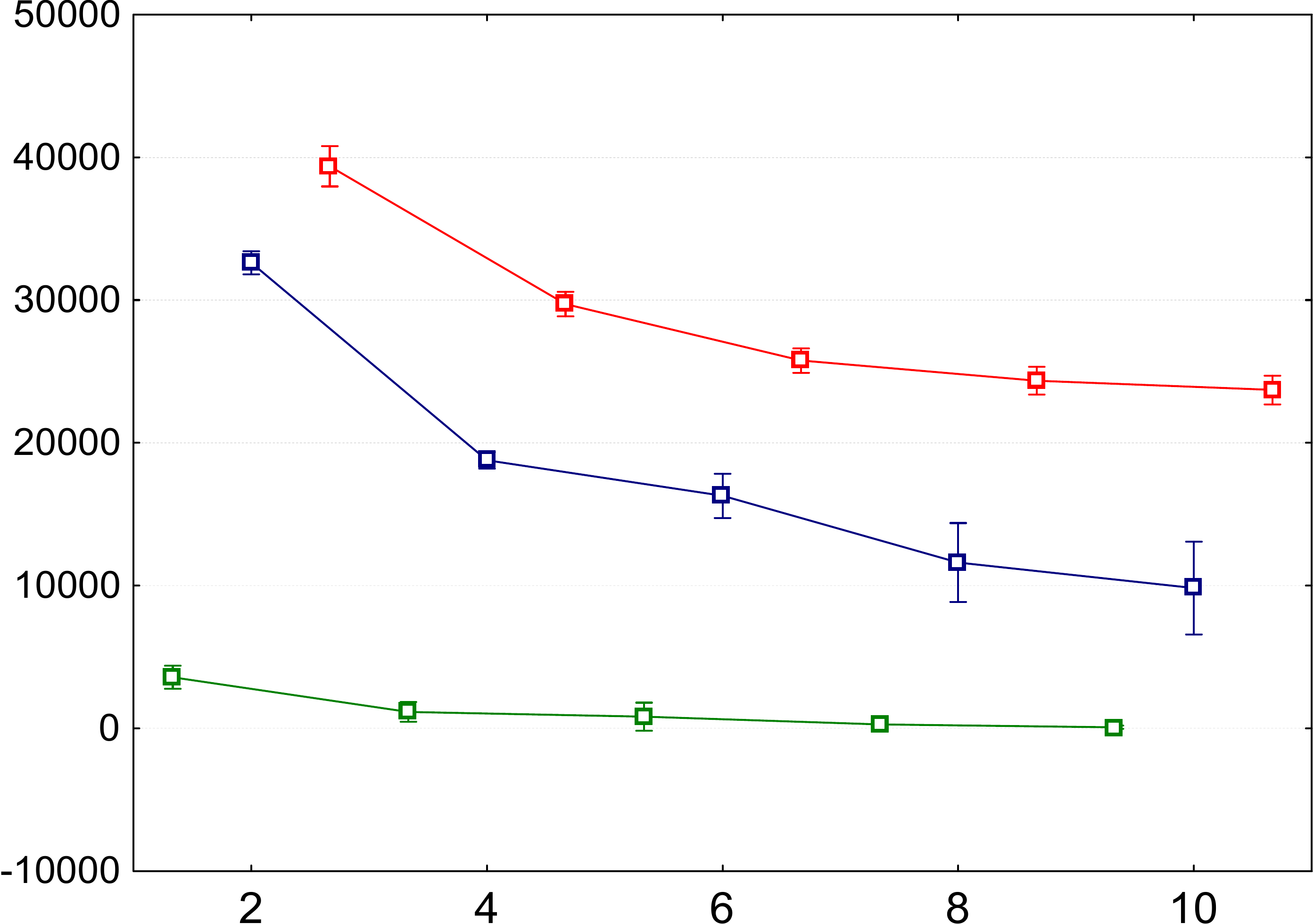}}
     &
\makecell{(h)\\\includegraphics[align=c,width=0.31\textwidth]{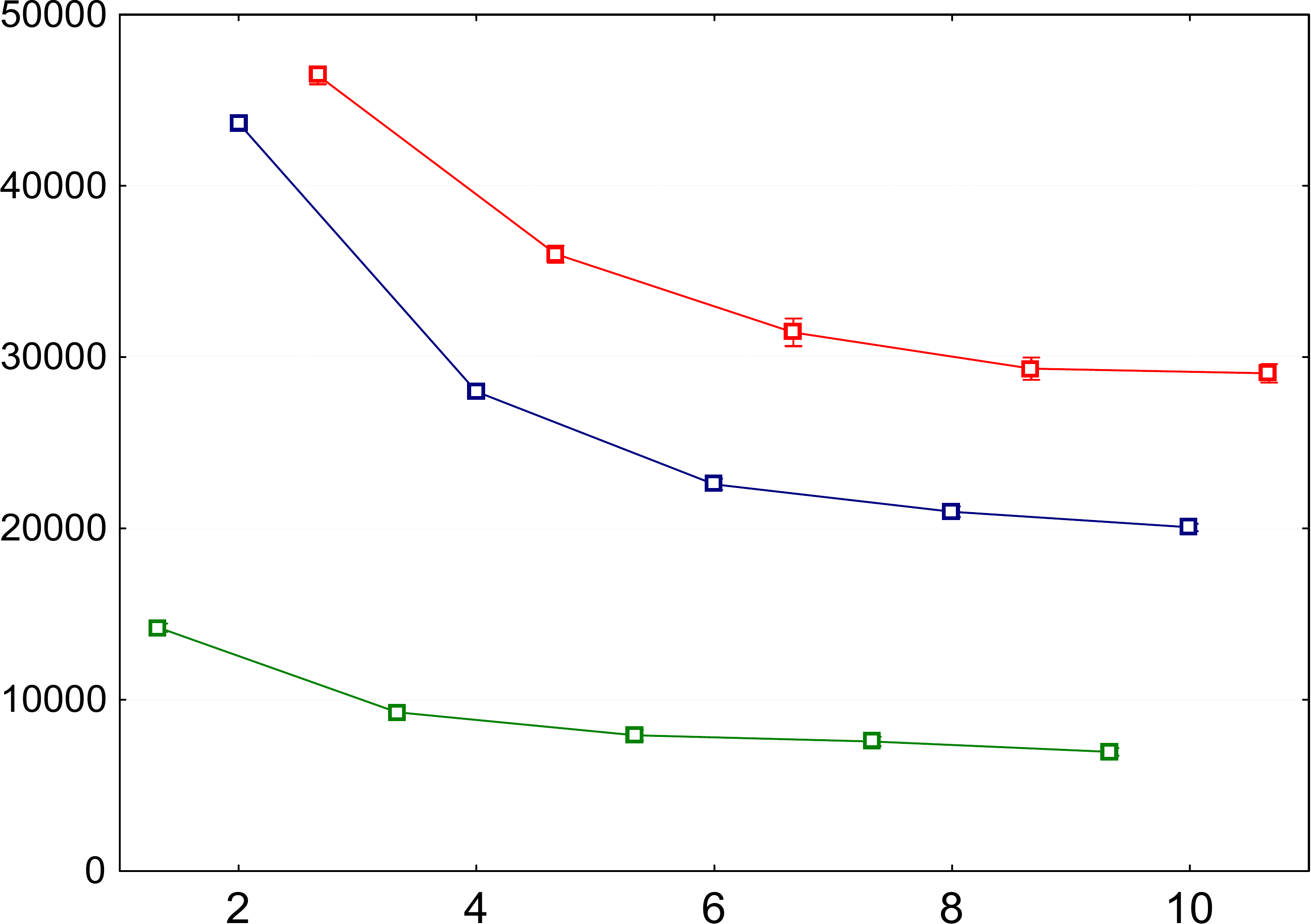}}
     &
\makecell{(i)\\\includegraphics[align=c,width=0.31\textwidth]{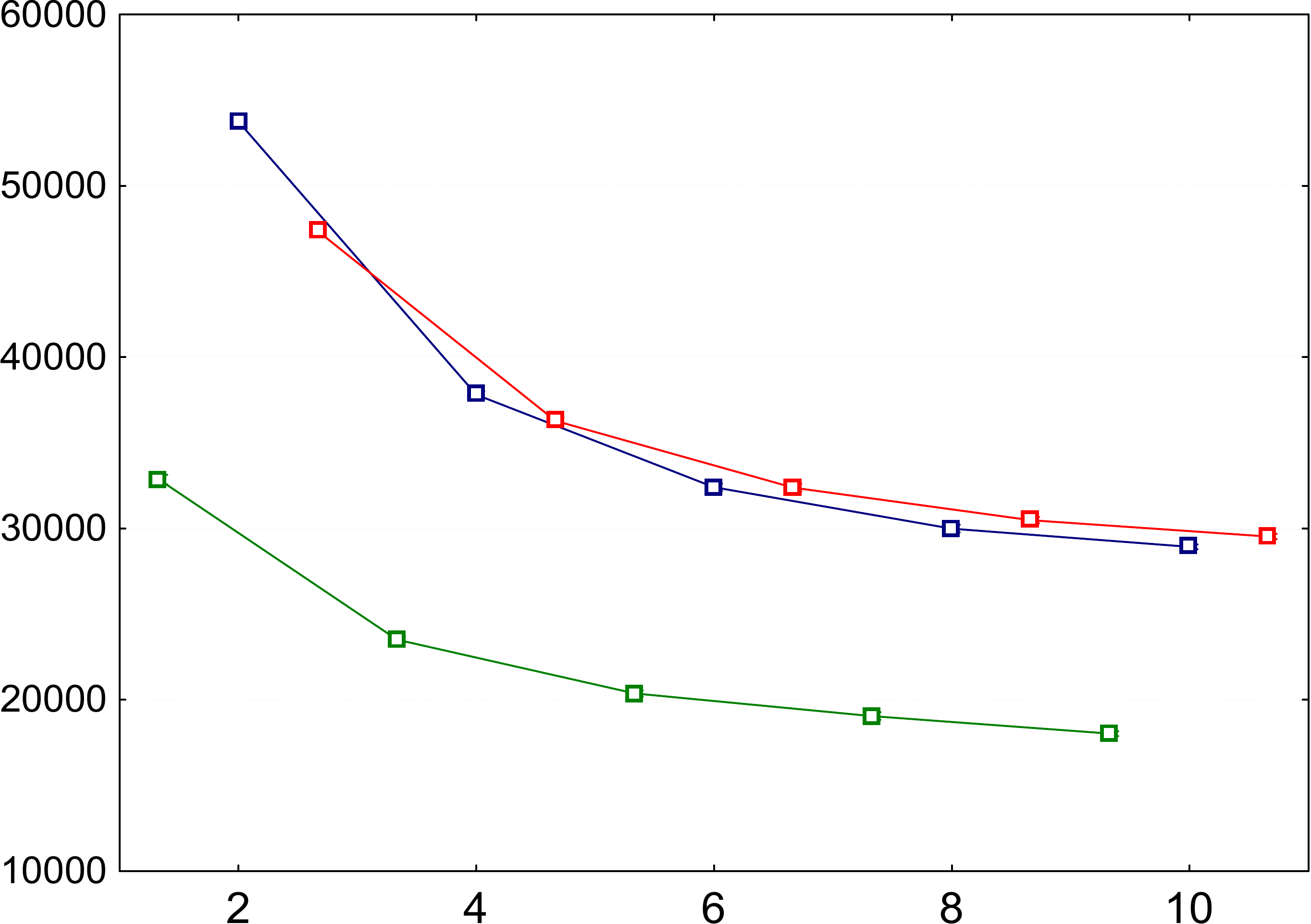}}\\
\rotatebox[origin=c]{90}{\footnotesize \quad\enskip cell quality ($\mathbf{Q}_A$)\quad\enskip } &
\makecell{(j)\\\includegraphics[align=c,width=0.3\textwidth]{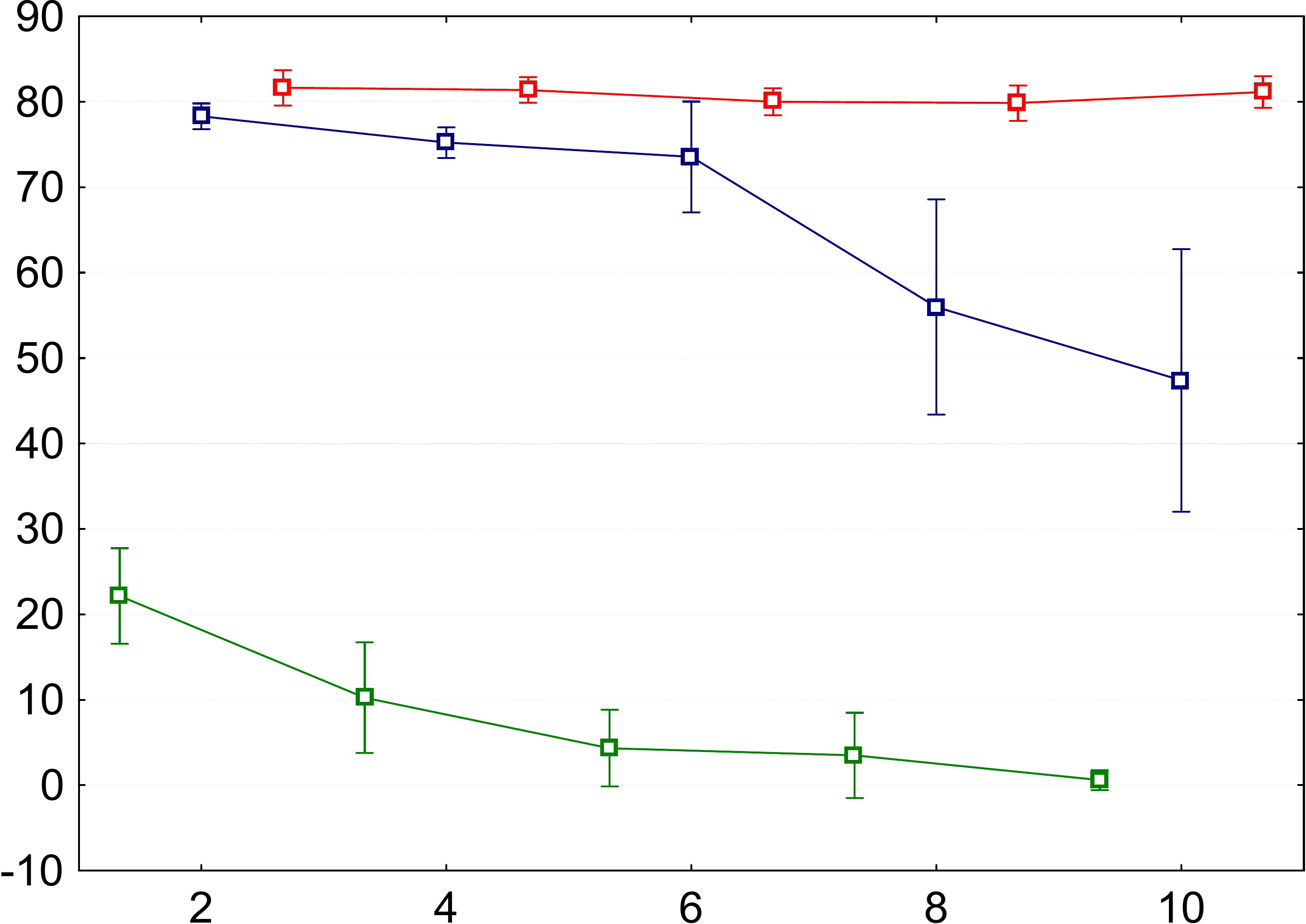}}
     &
\makecell{(k)\\\includegraphics[align=c,width=0.3\textwidth]{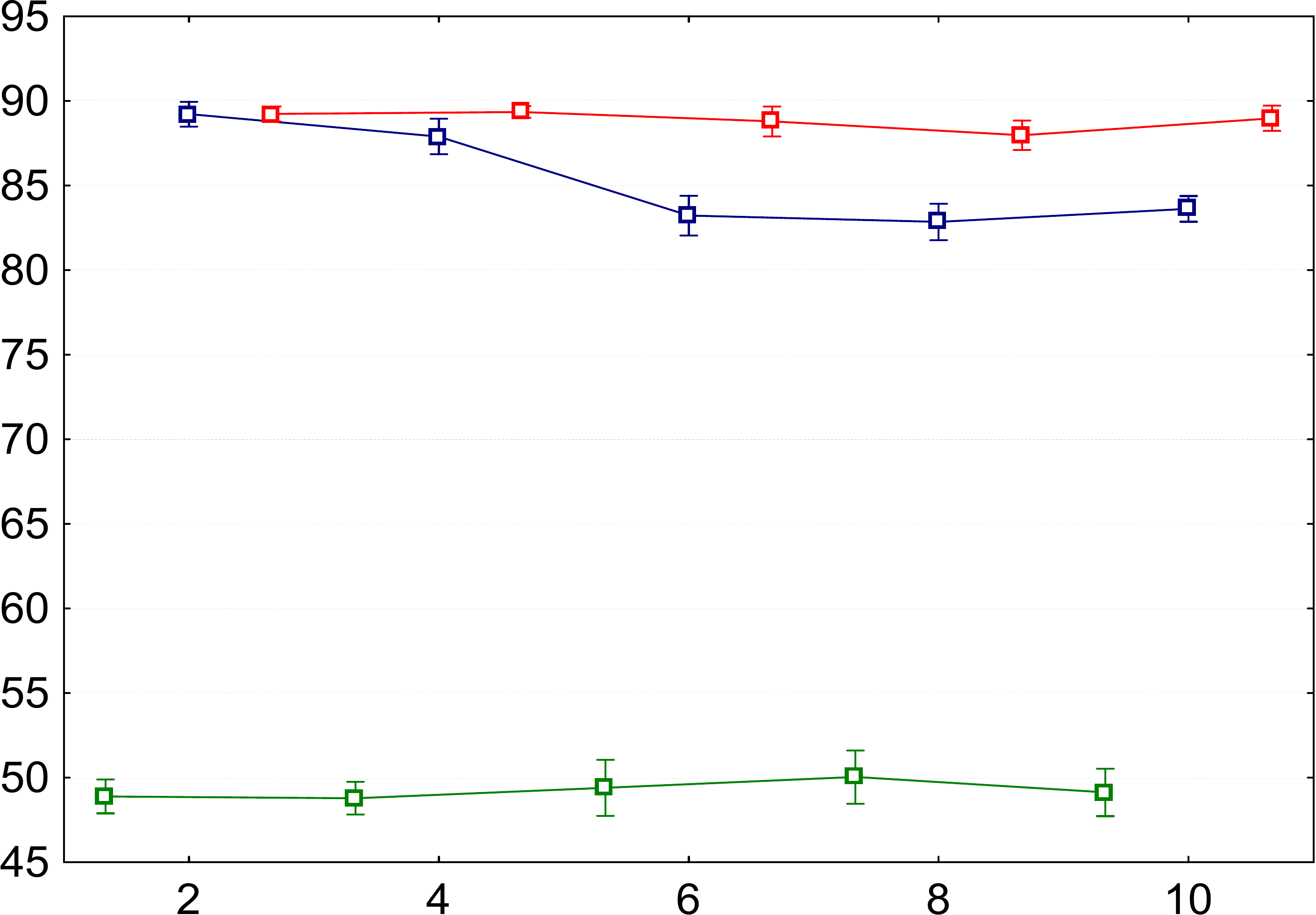}}
     &
\makecell{(l)\\\includegraphics[align=c,width=0.3\textwidth]{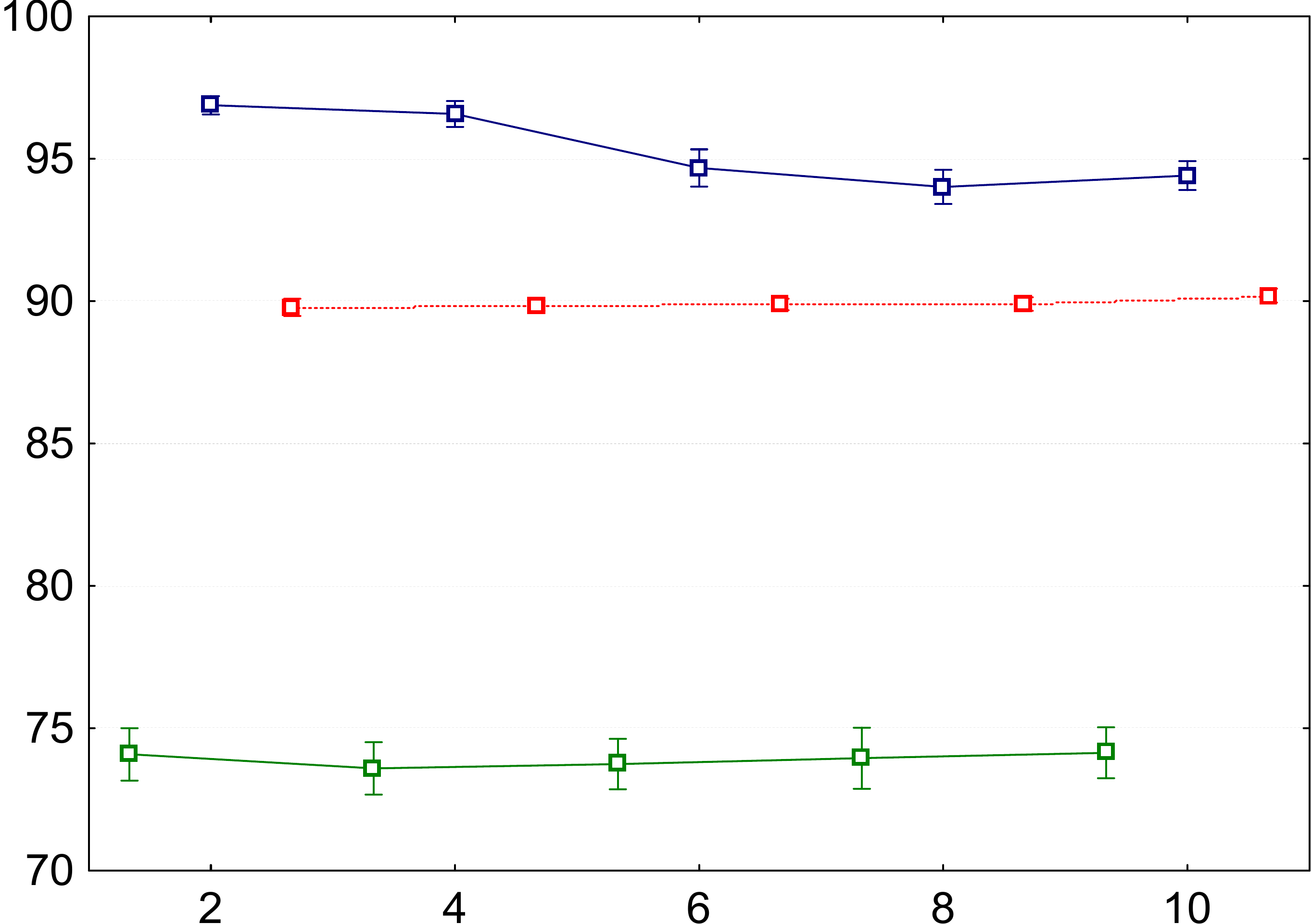}}
\end{tabular}
    \caption{Performance of different search strategies (average and its standard deviation in 30 experiments) along average metrics in relation to the proportions of bad, medium and good habitat (``map type'') and winter harshness. Red, green and blue lines represent respectively ``Smart'', ``Random'' and ``Dreamer'' strategies. X-axis indicates winter harshness, Y-axis - the values of the respective metrics (see \eqref{b_aver_n}\eqref{b_aver_all} for the explicit expressions of the shown metrics)}
    \label{fig:2d_aver}
\end{figure}
\endgroup

\newpage

\begingroup
\setlength{\tabcolsep}{5pt}

\begin{figure}[H]
    \centering
\begin{tabular}{cccc}
& {\footnotesize``bad'' map} & {\footnotesize``medium'' map} & {\footnotesize``good'' map} \\
\rotatebox[origin=c]{90}{\footnotesize \qquad survival ($\mathbf{S}$)\qquad} & \makecell{(a)\\
\includegraphics[align=c,width=0.3\textwidth]{pics/av_survival_bad1.pdf}}
     &
\makecell{(b)\\\includegraphics[align=c,width=0.3\textwidth]{pics/av_survival_medium1.pdf}}     
     &
\makecell{(c)\\\includegraphics[align=c,width=0.3\textwidth]{pics/av_survival_good1.pdf}}\\
\rotatebox[origin=c]{90}{\footnotesize \qquad dispersal ($\mathbf{D}_S$)\qquad} &
\makecell{(d)\\\includegraphics[align=c,width=0.3\textwidth]{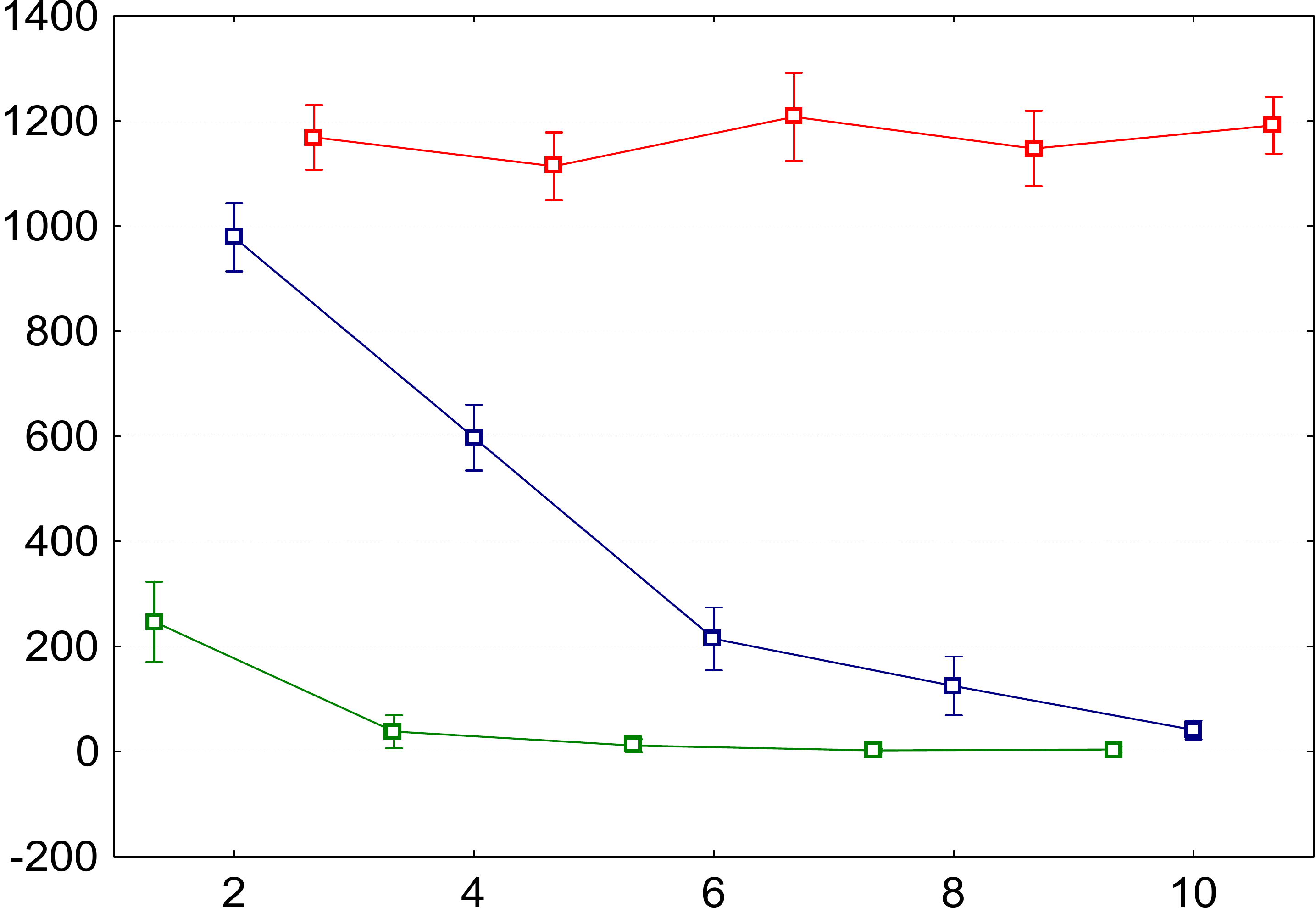}}
     &
\makecell{(e)\\\includegraphics[align=c,width=0.3\textwidth]{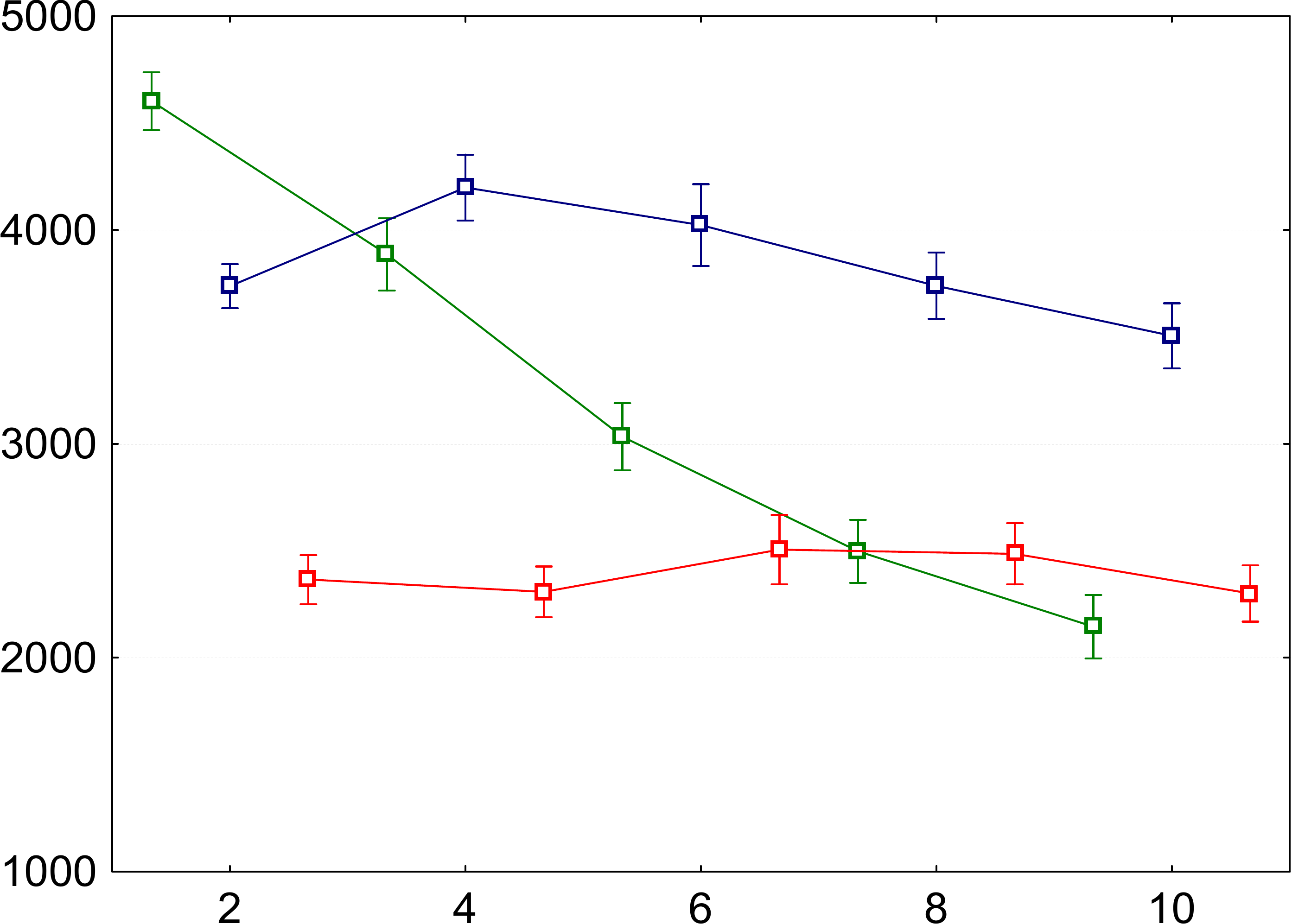}}     
     &
\makecell{(f)\\\includegraphics[align=c,width=0.3\textwidth]{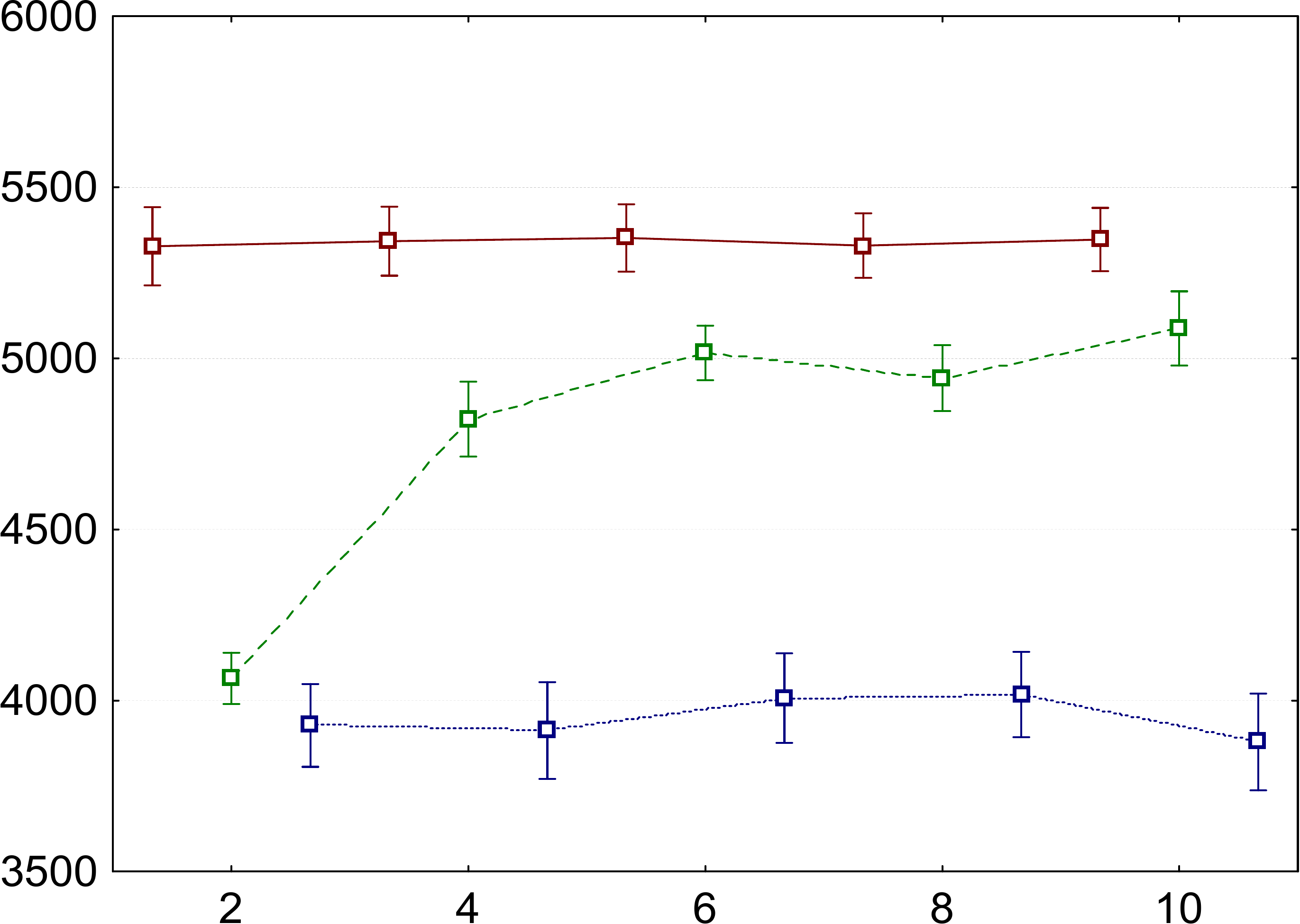}}\\
\rotatebox[origin=c]{90}{\footnotesize \qquad energy ($\mathbf{E}_S$)\qquad} &
\makecell{(g)\\\includegraphics[align=c,width=0.3\textwidth]{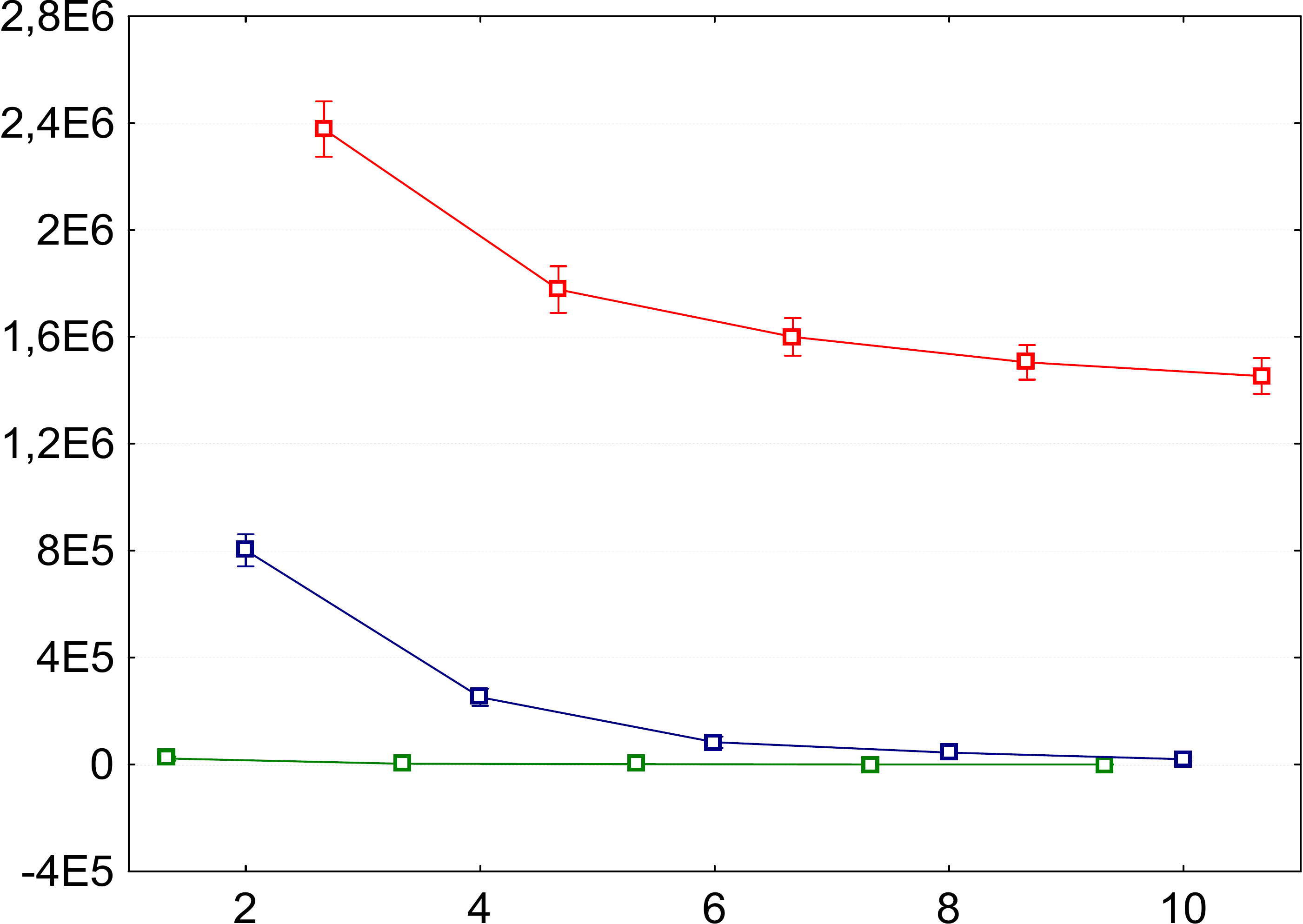}}
     &
\makecell{(h)\\\includegraphics[align=c,width=0.3\textwidth]{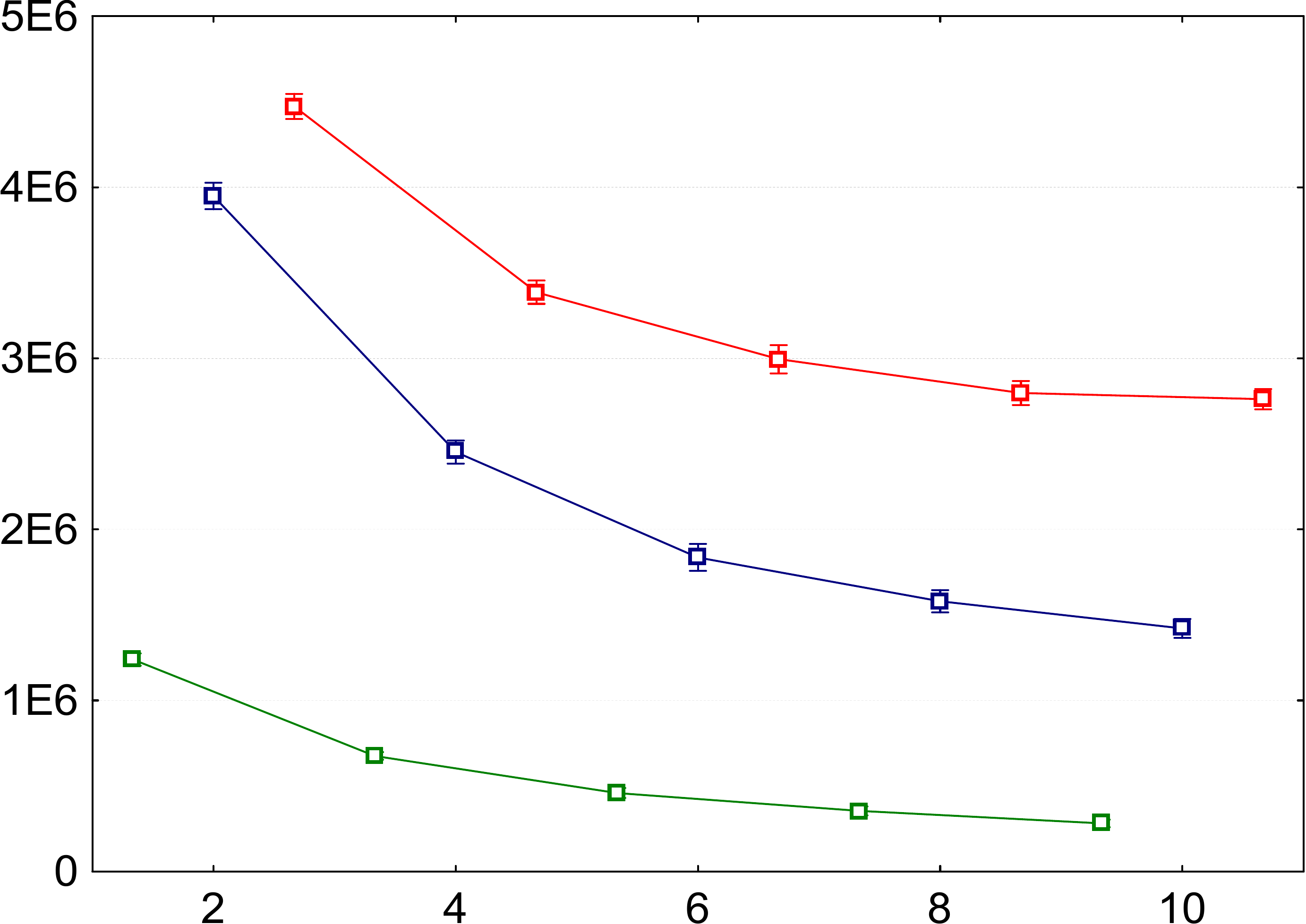}}     
     &
\makecell{(i)\\\includegraphics[align=c,width=0.3\textwidth]{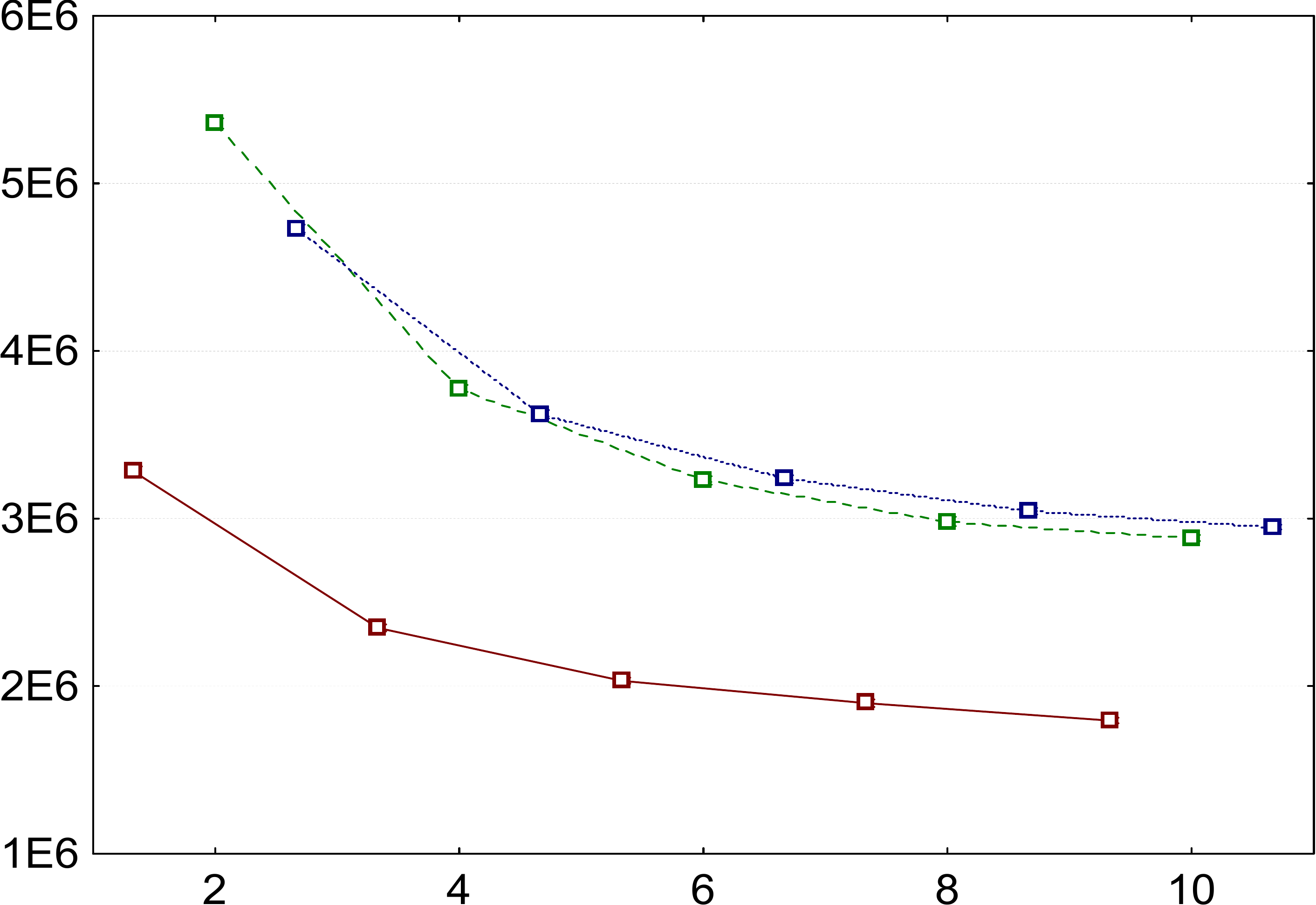}}\\
\rotatebox[origin=c]{90}{\footnotesize \quad\enskip cell quality ($\mathbf{Q}_S$)\quad\enskip } &
\makecell{(j)\\\includegraphics[align=c,width=0.3\textwidth]{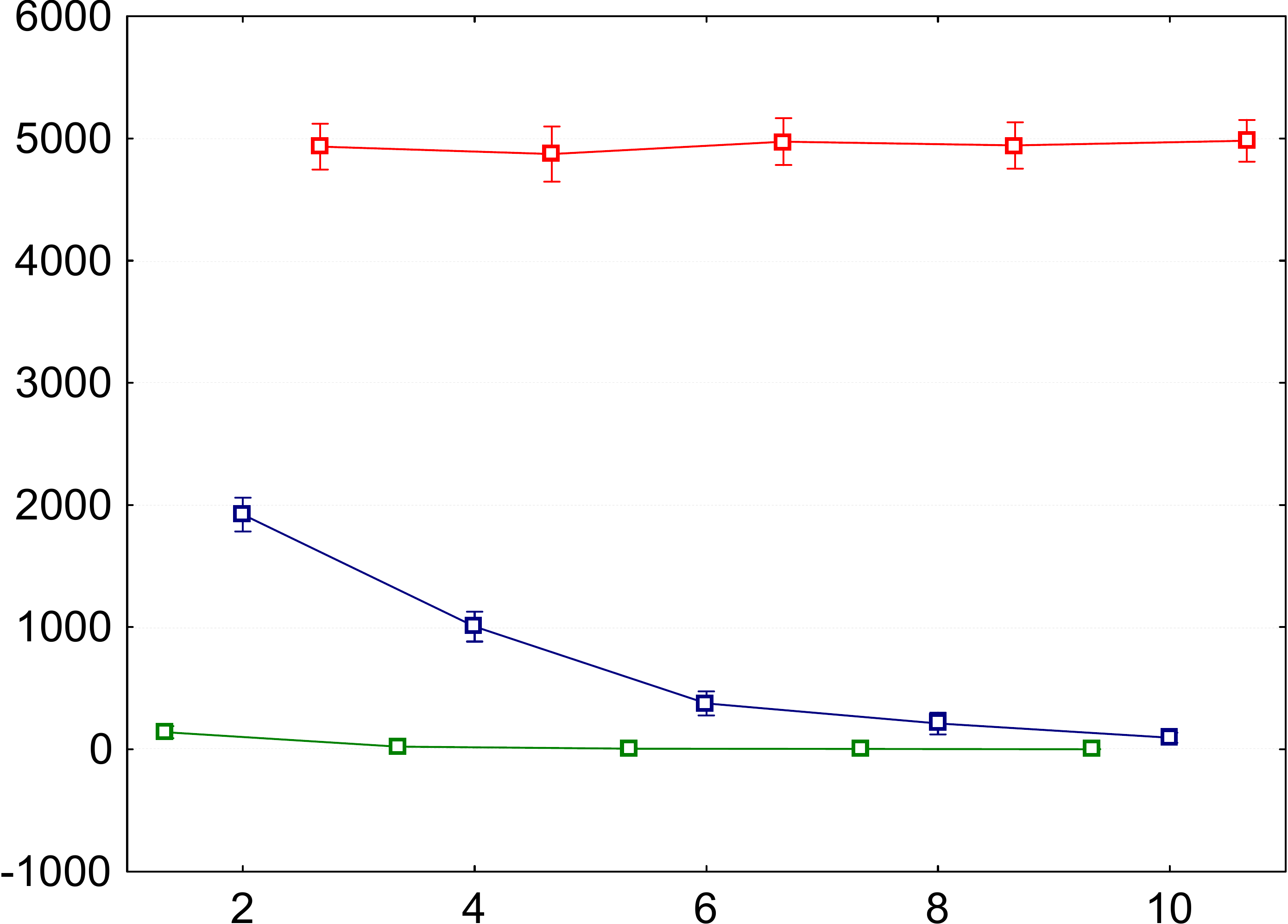}}
     &
\makecell{(k)\\\includegraphics[align=c,width=0.3\textwidth]{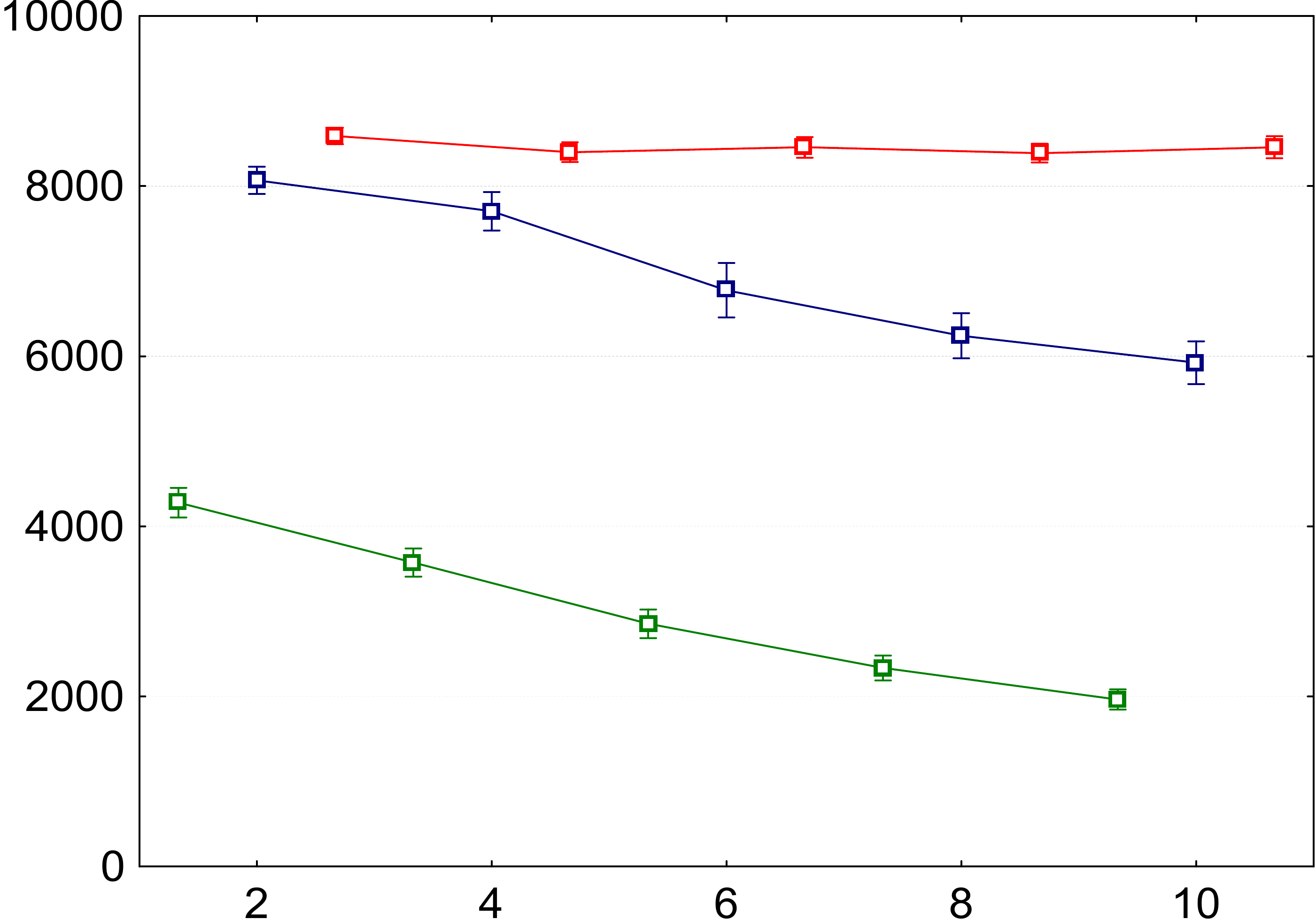}}     
     &
\makecell{(l)\\\includegraphics[align=c,width=0.3\textwidth]{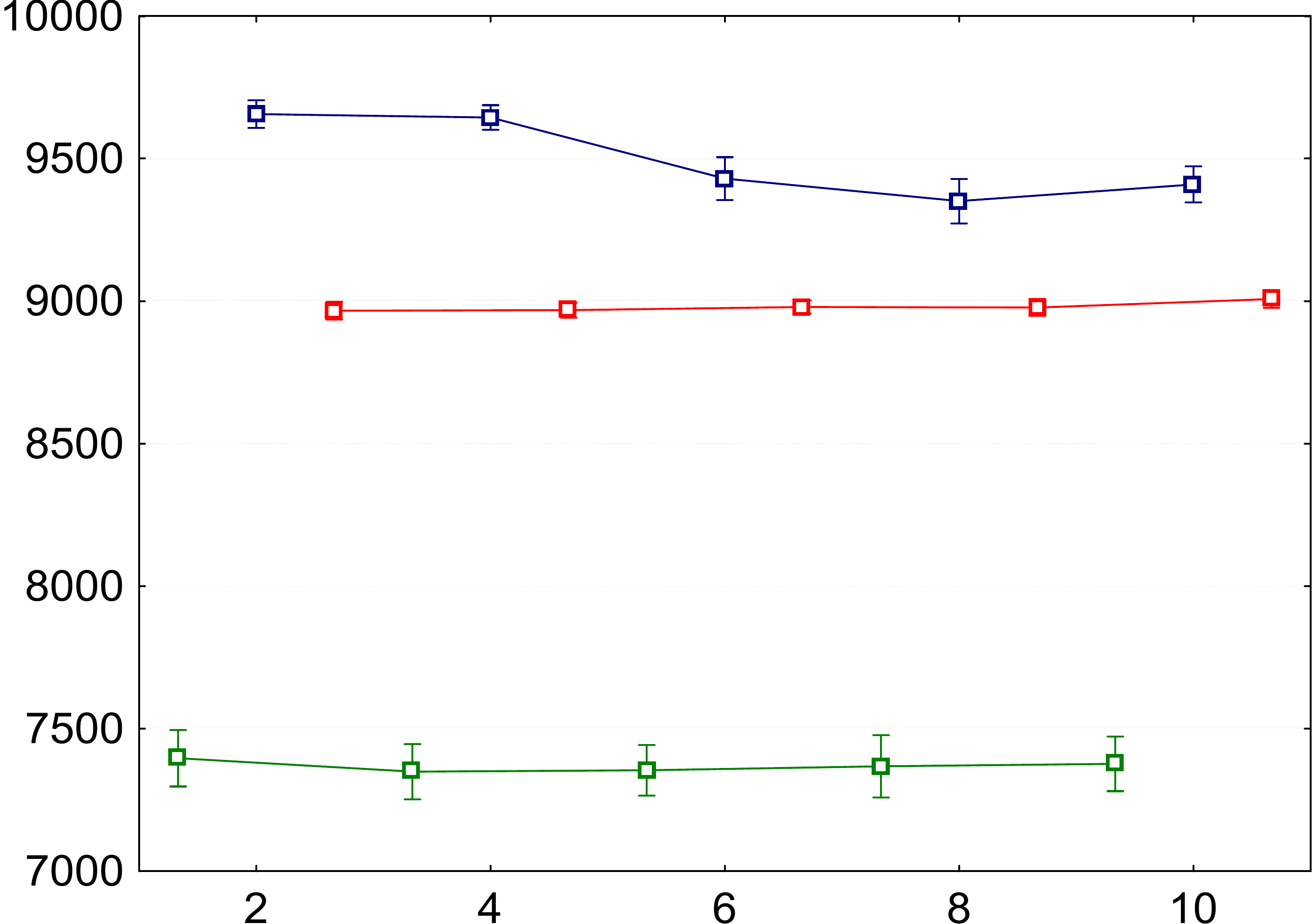}}
\end{tabular}
    \caption{Performance of different search strategies along summary metrics (sum and its standard deviation in 30 experiments)in relation to the proportions of bad, medium and good habitat (``map type'') and winter harshness. Red, green and blue lines represent respectively ``Smart'', ``Random'' and ``Dreamer'' strategies. X-axis indicates winter harshness, Y-axis -- the values of the respective metrics (see \eqref{b_aver_n}\eqref{eq:b_sum_all} for the explicit expressions of the shown metrics)}
    \label{fig:2d_sum}
\end{figure}

\endgroup


\subsection{Multivariate portrays of the strategies}

In Sections 3.1 - 3.3 we demonstrated that the relative success of a strategy may consist of different components: e.g., one strategy may win in dispersal, another -- in energy or cell quality. To compare the relative performance of the considered search strategies along different metrics we use radar charts (Fig.\ref{fig:averaged_radars}) with all the metrics normalized into [0,1] (see equations \eqref{eq:normalized1},\eqref{eq:normalized2} in \ref{sec:algorithms}). For the detailed portrays considering the winter harshness see Fig.\ref{fig:radar_aver},\ref{fig:radar_sum} in \ref{sec:pictures}.


\begingroup
\setlength{\tabcolsep}{10pt} 
\begin{figure}[H]
    \centering
\begin{tabular}{cccc}
\qquad & {\footnotesize``bad'' map} & {\footnotesize``medium'' map} & {\footnotesize``good'' map} \\
\rotatebox[origin=c]{90}{\footnotesize \quad\qquad Summary\qquad\quad} &
\scalebox{0.7}{\raisebox{-.5\height}{\footnotesize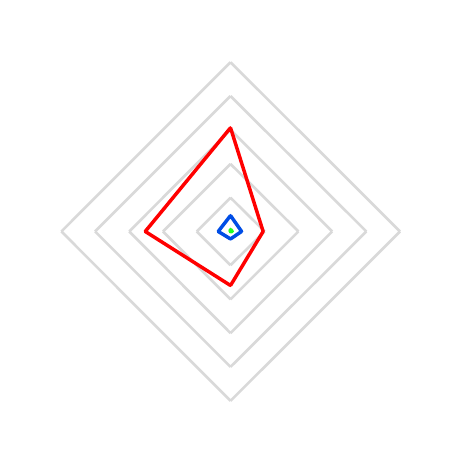}}
     &
\scalebox{0.7}{\raisebox{-.5\height}{\footnotesize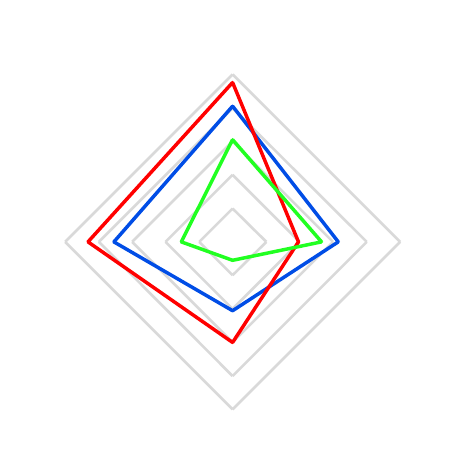}}
     &
\scalebox{0.7}{\raisebox{-.5\height}{\footnotesize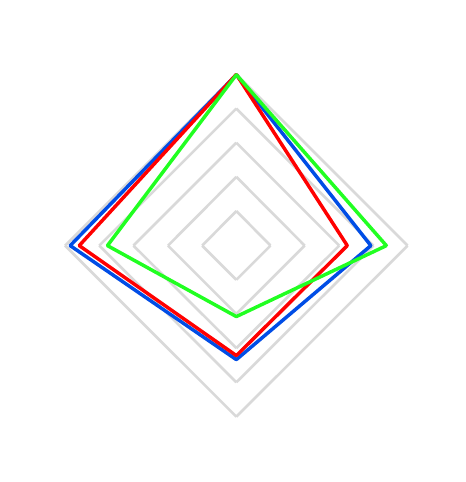}}
\\
\rotatebox[origin=c]{90}{\footnotesize \quad\qquad Average \qquad\quad} &
\scalebox{0.7}{\raisebox{-.5\height}{\footnotesize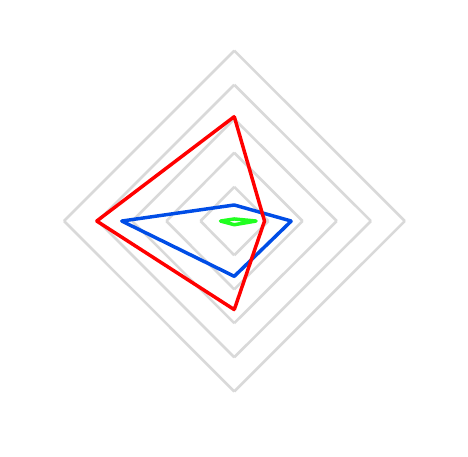}}
     &
\scalebox{0.7}{\raisebox{-.5\height}{\footnotesize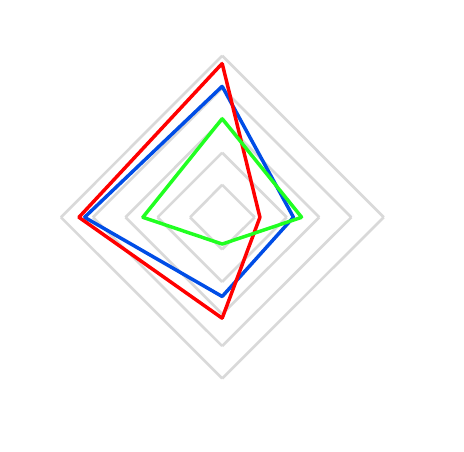}}
     &
\scalebox{0.7}{\raisebox{-.5\height}{\footnotesize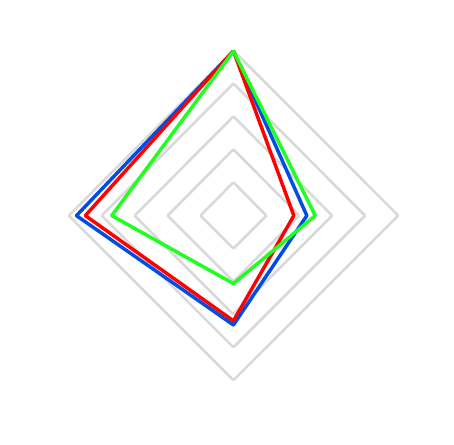}}

\end{tabular}
    \caption{Portrays of relative performance of the search strategies along different normalized metrics 
    averaged by the winter harshness $w$. 
    Color is used to mark off the strategies: green for ``Random'', red for ``Smart'' and blue for ``Dreamer''. All the axes run from 0 to 1 }
    \label{fig:averaged_radars}
\end{figure}

\endgroup

In Fig.\ref{fig:pareto} we compare different strategies along two general dimensions of success -- {\it Dispersal} D and {\it Fitness} EQ (see \ref{sec:metrics_general}) -- with the survival rate S playing the role of ``significance'' for each entry (all the metrics are again normalized into [0,1], see \eqref{eq:normalized1},\eqref{eq:normalized2} in \ref{sec:algorithms}). In all the cases, except for the summary metrics on ``bad'' maps, ``Dreamers'' clearly belong to the Pareto frontier (\cite{pareto}) (so that no other strategy could outperform it in both D and EQ). 


\begingroup
\setlength{\tabcolsep}{10pt} 

\begin{figure}[H]
    \centering
\begin{tabular}{ccc}
\qquad & Summary &  Average \\ 
\rotatebox[origin=c]{90}{\qquad\qquad``bad'' map\qquad\qquad} &
\makecell{(a)\\\scalebox{1.2}{\raisebox{-.5\height}{\footnotesize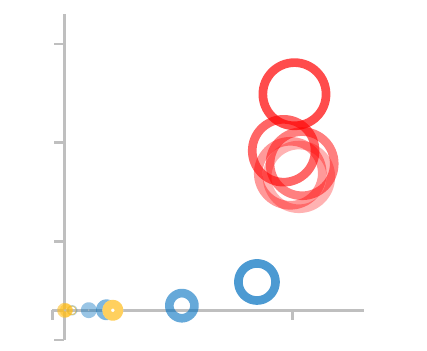}}}
     &
\makecell{(b)\\\scalebox{1.2}{\raisebox{-.5\height}{\footnotesize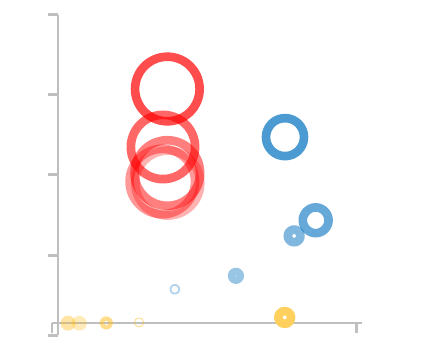}}}
\\
\rotatebox[origin=c]{90}{\quad\qquad``medium'' map\qquad\quad} &
\makecell{(c)\\\scalebox{1.2}{\raisebox{-.5\height}{\footnotesize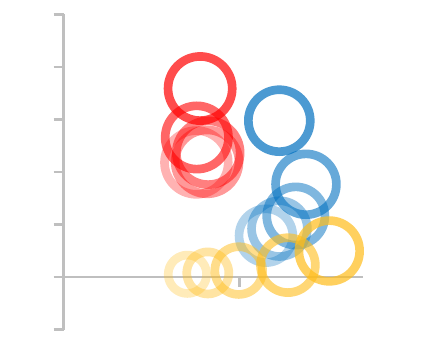}}}
     &
\makecell{(d)\\\scalebox{1.2}{\raisebox{-.5\height}{\footnotesize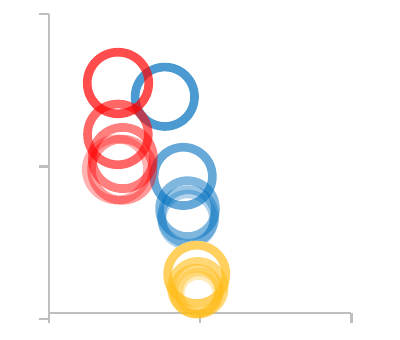}}}
\\
\rotatebox[origin=c]{90}{\qquad\qquad``good'' map\qquad\qquad} &
\makecell{(e)\\\scalebox{1.2}{\raisebox{-.5\height}{\footnotesize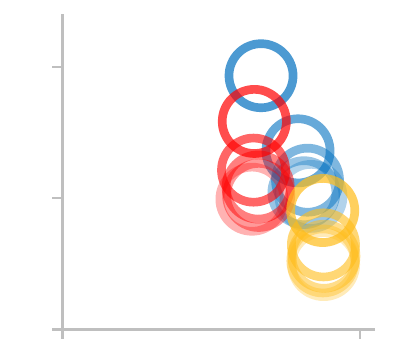}}}
     &
\makecell{(f)\\\scalebox{1.2}{\raisebox{-.5\height}{\footnotesize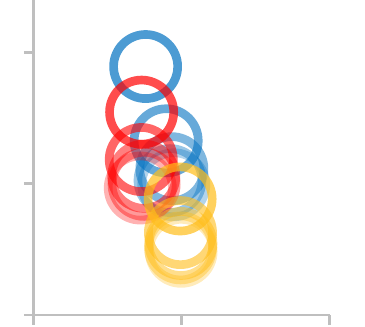}}}

\end{tabular}
    \vspace{1mm}
    \caption{Success of different strategies in {\it Dispersal} $\times$ {\it Fitness} space. Here red color corresponds to ``Smarts'', blue -- to ``Dreamers'' and orange -- to ``Randoms''. Color intensity indicates the harshness of winter $w$: the higher intensity -- the milder winter. The size of the circle is directly proportional to the corresponding survival rate $\mathbf{S}$. All the axes run from 0 to 1 (see \eqref{eq:normalized1}\eqref{eq:normalized2} for the details)}
    \label{fig:pareto}
\end{figure}

\endgroup

\section{Discussion}

As mentioned in \cite{Bowler2005Causes}, how the organism integrates the information from different environmental cues and thereby reaches its decision has largely been overlooked in previous studies and this makes it difficult to compare our results with other models of animals’ dispersal.

For example, in differ with the studies that question the dynamics of dispersing as a function of some external factors (like habitat quality, habitat matrix, climate, and other factors that affect dispersal costs) (\cite{Kubisch2014Where}), we suggest that dispersal success depends on the “built-in” behavioral strategy which could be more or less successful according to the particular external factors. In differ with the studies that also base on some behavioral strategy, e.g. implying a-priori differences between short-distance and long-distance dispersers (\cite{Ramanantoanina2014Spatial}), the probability to perform a long movement is equal for all the animats and drawn from the Levy-walk model.  The habitat preferences are in fact also the same for all animats (as could be expected for the individuals belonging to one species) and base on the potential energy gain from the cell.   Thus the model populations are different rather in behavioral than ecological traits and their success in survival is affected by the decision-making process. The strategies (``Random'', ``Smart'', ``Dreamer'') are the examples of search strategies previously implemented in a number of studies (e.g. \cite{Stamps2005Search, Zollner1999Search}), but they have not been compared. The classical study \cite{Zollner1999Search} also compares the success of dispersal in habitat of low, medium and high habitat quality, but they address differences related rather to the shape of movements than to decision-making process underlying these movements. In our study we do not pre-define the shape of movements though all the strategies perform the correlated Levy walk.

Our model is pretty simple: it does not take into account possible interactions between population members and habitat depletion. We assume that animals are released in a region previously non-inhabited by the species. In such new environment both intra-species competition (and other types of intra-population interactions) is negligible and the resources of the habitat do not limit animals’ abundance. An important part of our model is that all the individuals following a particular strategy are exactly the same, while animals' personality is an important factor shaping their movements (\cite{Hawkes2009Linking}). We believe that more complex models that take into account the variability of movement behavior could be the subject of future research.

``Random'' is a strategy with no habitat selection. It could be implemented when animals cannot correctly assess habitat quality. In reality, errors in the estimation of habitat quality could result in ``ecological trap'' (\cite{Kokko2001Ecological}) while random search theoretically could help to escape these traps and provide higher fitness in compare with more specific strategies of habitat selection. However such scenario was not included in our model: ``Smarts'' and ``Dreamers'' correctly assess the habitat quality (at the strata level -- ``good'', ``medium'' or ``bad'') and thus random search predictably performed worse than the other strategies (in terms of survival, accumulated energy and the quality of habitat) in marginal (``bad'') and suboptimal (``medium'') environment. This is in agreement with other studies that compare random search with different search strategies (\cite{Conradt2003Foray}). Interestingly, in favorable environment (``good'') survival and fitness of ``Randoms'' stays as high as in the other strategies, while being accompanied with the highest dispersal ability. Thus random search strategy could explain very fast expansion and establishment of viable populations in good environment (for example, when species relax the effect of competitors, parasites and diseases) which is the classic case of biological invasion (\cite{Lewis2015Mathematics} and references therein).

``Smarts'' are the best dispersers only in bad habitat and mainly due to low survival of ``Randoms'' and ``Dreamers''. In medium and good habitat the ``Smart'' strategy provides high survival and high fitness combined with relatively low dispersal distances -- organisms sequentially inhabit the best patches close to the place of origin. We suspect that this strategy could underlie the wave front expansion (type I expansion in terms of \cite{Shigesada2002Invasion}) which is slow expansion with constant speed. Though the speed of population expansion in our model has  not yet been studied, the independence of dispersal from winter conditions allows to assume the independence from temporal climatic variation and thus to support the suggestion of constancy of the expansion rate. Fast discovery of locally optimal habitat, demonstrated by ``Smarts'', could suggest the familiarity of the dispersers with the habitat, thus it could be expected for the populations introduced in the environment which is ecologically close to that in the place of origin (\cite{Davis2004effect}). A practical way to make animals familiar to a new environment during re-introduction is keeping them in enclosures for some time before release. It is well known by conservation biologists and called `` soft release'' (\cite{Jung2017Observations}).

``Dreamers'' settled the best possible habitat in the middle and far distance from the place of release, demonstrating high survival and high energy accumulation by dispersing individuals. Similar effect has been previously reported in \cite{Stamps2005Search} where the likelihood that a disperser would settle in a high-quality rather than in a low-quality habitat is positively related to both selectivity of habitat and time of search. Dreamers are indeed highly selective due to high difference between the available habitat and the ``habitat of dream'' and this strategy could illustrate the dispersal associated with the habitat selection based on natal habitat preference induction (NHPI) (\cite{Davis2004effect,Stamps2007Someplace}). Extending the result of \cite{Stamps2005Search}, we show how this pattern is affected by habitat composition, seasonal variation in habitat quality and compare it with the success of the other strategies. Our results indicate that ``Dreamers'' are quite successful in finding good habitat at a significant distance from the place of origin, especially on ``medium'' and ``good'' maps, though the success of dispersal remains relatively low in bad habitat depending in addition on winter harshness. Populations implementing the ``Dreamer'' strategy thus expands with successfully dispersing propagules rather than with a wave front.

In the above paragraphs we tried to assign the results of our experiments to the real ecological situations and types of population expansion. However, in reality it is most probable that the expansion of a population results from a mixed strategy, where different individuals use different strategies. In the studies of animals’ personality, researchers most often distinguish ``bold'' and ``shy'' individuals, but more specifically to dispersal they are placed in the ``exploration-avoidance'' continuum, illustrating the animals’ reaction to novel environment (\cite{Reale2007Integrating}). Populations with high dispersal rates have been shown to be composed of bold, exploratory, and aggressive individuals, which is being found in different taxons and habitat (\cite{Malange2016Personality} and references therein). These better dispersers in terms of our model could be assigned to the ``Random'' and ``Dreamer'' strategies. High mortality, which is often observed for bold and explorative individuals (\cite{Bremner-Harrison2004Behavioural}), put them closer to ``Randoms'' than to ``Dreamers'', since the latter are those who used their high dispersal ability to successfully settle an optimal habitat. `Shy'' individuals could follow the ``Smart'' strategy.

\section{Conclusions and future work}

Answering the questions formulated in the Introduction we conclude that

1) In good and medium habitat the ``Dreamer'' strategy performs better than the others since it provides higher dispersal in compare with the ``Smart'' strategy and higher quality of habitat in the place of settlement in compare with the ``Random'' strategy. In bad habitat only ``Smarts'' provides a reasonable level of survival and population fitness. The ``Random'' strategy is appropriate only in good habitat;

2) The performance of the “Smart” strategy does not depend strongly on winter harshness, while for the two other strategies most metrics decrease with winter harshness. The only metrics that clearly increased with winter harshness is the dispersal rate of ``Dreamers''in good habitat;

3) The ``Dreamer'' strategy or the strategy of deferred gain belongs to the Pareto frontier in the {\it Fitness$\times$Dispersal} space but only in optimal and suboptimal habitat and in the relatively mild climate.

The most general conclusion is that while survival and wealth of the population is affected presumably by overall habitat quality, the dispersal depends mainly on the behavioral strategy. In the future work, it would be interesting to model the population expansion for several generations checking the scenarios of random and non-random inheritance of the search strategies.

\section*{Author's contributions} 

N.I. Markov: Conceptualization, Methodology, Formal analysis, Investigation, Writing - Review \& Editing; E.E. Ivanko: Methodology, Software, Formal analysis, Investigation, Resources, Data Curation, Writing -- Original Draft, Visualization, Funding acquisition.

\section*{Funding information}

The research was supported by Russian Foundation for Basic Research (RFBR), project No 18-07-00637а.

\appendix

\section{Formal algorithms and metrics}
\label{sec:algorithms}

\subsection{Map generation}
\label{alg_sec_map}
Each computation experiment modelling the population dispersal starts from the following simple algorithm which generates a new map in accordance with certain parameters.

\vspace{10pt}

\begin{samepage}
\noindent{\bf Algorithm 1}: $L\times L$ map generation
\begin{enumerate}
    \item Choose the size $L$ of the map and the numbers of the cells of each strata type: $S_1$ of quality 1 (``bad''), $S_2$ of quality 2 (``medium'') and $S_3$ of quality 3 (``good'') so that $S_1+S_2+S_3=L\times L$;
    \item Choose the limits of the cells energy for each type (quality) of strata: $\mathcal{E}_{min}^{1}<\mathcal{E}_{max}^{1}<\mathcal{E}_{min}^{2}< \mathcal{E}_{max}^{2}<\mathcal{E}_{min}^{3}<\mathcal{E}_{max}^{3}$;
    \item Let initially all the cells be ``medium'': $\forall x,y\in\overline{1,L}\times\overline{1,L}\ \ \mathcal{M}(x,y):=2$ (strata type function);
    \item Choose the number of ``brushes'' $P\in\overline{1,10}$ (this parameter affects the ``patchness'' of the resulting map) and two brush sizes $B_1, B_2\in\overline{1,5}$;
    \item Move $P$ brushes randomly and independently (random walk); each brush centered at the cell $(x_0,y_0)$ paints all the cells $(x,y)$ such that $\max\{|x-x_0|,|y-y_0|\}\le B_1$ in color 1 -- ``bad'', i.e. $\mathcal{M}(x,y):=1$; continue this painting until the painted area is equal to $S_1+S_3$; 
    \item Move $P$ brushes randomly and independently (random walk); each brush centered at the cell $(x_0,y_0)$ paints all the cells $(x,y)$ of color 1 (i.e. $\mathcal{M}(x,y)=1$) such that $\max\{|x-x_0|,|y-y_0|\}\le B_2$ in color 3 -- ``good'', i.e. $\mathcal{M}(x,y):=3$; continue this painting until the painted area is equal to $S_3$;
    \item Choose the energy $\mathcal{E}(x,y)$ of each cell $(x,y)$ randomly: ``bad'' -- from $\mathcal{E}_{min}^{1}$ to $\mathcal{E}_{max}^{1}$; ``medium'' -- from $\mathcal{E}_{min}^{2}$ to $\mathcal{E}_{max}^{2}$ and ``good'' -- from $\mathcal{E}_{min}^{3}$ to $\mathcal{E}_{max}^{3}$;
    \item Return $\mathcal{M}\colon L\times L\rightarrow \overline{1,3}$ and $\mathcal{E}\colon L\times L\rightarrow \overline{\mathcal{E}_{min}^{1},\mathcal{E}_{max}^{3}}$.
\end{enumerate}
\end{samepage}

In our experiments we used $L=100$, $P=5$, $B_1=B_2=1$, $\mathcal{E}_{min}^{1}=0$, $\mathcal{E}_{max}^{1}=20$, $\mathcal{E}_{min}^{2}=40$, $\mathcal{E}_{max}^{2}=60$, $\mathcal{E}_{min}^{3}=80$, $\mathcal{E}_{max}^{3}=100$ and three types of map: 1) ``bad'' $6700/3000/300$; 2) ``medium'' $2200/5600/2200$ and 3) ``good'' $300/3000/6700$. 

\subsection{Population metrics}
\label{sec:b_metrics}
Before proceeding to the algorithms expressing different dispersal strategies let us introduce the metrics we used to evaluate the condition of the populations. Each dispersal modelling experiment was repeated $J$ times for each possible tuple $(m,u,w)$, where
 $m$ is the type of map (``bad'', ``medium'', ``good''), 
 $u$ -- dispersal strategy or the type of determination by environment  (``Random'', ``Smart'', ``Dreamer'') and $w$ -- winter harshness (2,4,6,8,10), so each of the following raw metrics is  by default parametrized with $(m,u,w)$. 
 The first raw metric is the population abundance: let $N_j$ be the number of alive animats at the final step of the $j$-th repetition (the starting number of animats $n^0$ is always 100). The other three raw metrics are specific to individual animats indexed by $k$. Note that if an animat is not alive at the end of a repetition, the corresponding raw metrics are set equal to zero. The second raw metric is the bias of each animat from its starting cell: 
 $$d_{jk}=\sqrt{(x_{jk}-x_{jk}^0)^2+(y_{jk}-y_{jk}^0)^2},$$
where $x_{jk},y_{jk}$ are the coordinates of the cell the $k$-th animat occupies at the final step of the $j$-th repetition and $x_{jk}^0,y_{jk}^0$ -- the initial coordinates of this animat. The third raw metric is the energy $e_{jk}$ accumulated by the $k$-th animat by the final step of the $j$-th repetition. And the last raw metric is the quality of habitat: $q_{jk}$ -- the energy (attractiveness) of the cell occupied by the $k$-th animat at the end of the $j$-th repetition. 

These raw metrics allow to construct several simple generalized characteristics of the population averaged by the repetitions and animats. 
The first group of metrics contains the number of survivors $\mathbf{N}$ and the survival rate $\mathbf{S}$:
\begin{equation}
\label{b_aver_n}
\mathbf{N}(m,u,w) = \frac{1}{J}\sum_{j=1}^J N_{j}(m,u,w),\ \ \mathbf{S}(m,u,w)=\mathbf{N}(m,u,w)/n^0.
\end{equation}
For other groups let us introduce auxiliary notations:
\begin{equation*}
\begin{aligned}
\widetilde{D}(m,u,w,j) &= \sum_{k=1}^{n^0} d_{jk}(m,u,w),\ \ \widetilde{E}(m,u,w,j) = \sum_{k=1}^{n^0}e_{jk}(m,u,w),\\
\widetilde{Q}(m,u,w,j) &= \sum_{k=1}^{n^0}q_{jk}(m,u,w),\ \ \widetilde{EQ}(m,u,w,j) = \sum_{k=1}^{n^0}q_{jk}(m,u,w)\, e_{jk}(m,u,w);
\end{aligned}
\end{equation*}
recall that $d_{jk}(m,u,w)=e_{jk}(m,u,w)=q_{jk}(m,u,w)=0$ if the $k$-th animat did not manage to survive until the end of the $j$-th repetition of the experiment with parameters $(m,u,w)$.

In these notations, the average metrics for dispersal distance $\mathbf{D}_A$, accumulated energy $\mathbf{E}_A$ and final cell quality $\mathbf{Q}_A$ can be expressed as:
\begin{equation}
\label{b_aver_all}
\begin{aligned}
\mathbf{D}_A(m,u,w) &= \frac{1}{J}\sum_{j=1}^J \tfrac{\widetilde{D}(m,u,w,j)}{N_{j}(m,u,w)},\ \ 
\mathbf{E}_A(m,u,w) = \frac{1}{J}\sum_{j=1}^J \tfrac{\widetilde{E}(m,u,w,j)}{N_{j}(m,u,w)},\\
\mathbf{Q}_A(m,u,w) &= \frac{1}{J}\sum_{j=1}^J  \tfrac{\widetilde{Q}(m,u,w,j)}{N_{j}(m,u,w)},\ \ 
\mathbf{EQ}_A(m,u,w) = \frac{1}{J}\sum_{j=1}^J \tfrac{\widetilde{EQ}(m,u,w,j)}{N_{j}(m,u,w)}.
\end{aligned}
\end{equation}

The last averaged metric $\mathbf{EQ}_A$ is an attempt to express the {\it fitness} of the survivors taking into account both the accumulated energy and the quality of occupied habitat. 

Averaged metrics characterize a single survivor. To address the conditions of the whole groups adopting different dispersal strategies we used the corresponding summary metrics (note that due to different survival rates in different repetitions the average and summary models may or may not be trivially dependent):
\begin{equation}
\footnotesize
\begin{aligned}
\mathbf{D}_S(m,u,w) &= \frac{1}{J}\sum_{j=1}^J \widetilde{D}(m,u,w,j),\ \ 
\mathbf{E}_S(m,u,w) = \frac{1}{J}\sum_{j=1}^J \widetilde{E}(m,u,w,j),\\
\mathbf{Q}_S(m,u,w) &= \frac{1}{J}\sum_{j=1}^J \widetilde{Q}(m,u,w,j),\ \ 
\mathbf{EQ}_S(m,u,w) = \frac{1}{J}\sum_{j=1}^J \widetilde{EQ}(m,u,w,j).
\end{aligned}
\label{eq:b_sum_all}
\end{equation}

Finally, let us normalize all the collected metrics values to be able to compare the input of each metric into the success of each strategy (see Fig.\ref{fig:averaged_radars}, \ref{fig:pareto},\ref{fig:radar_aver},\ref{fig:radar_sum}):
\begin{equation}
\footnotesize
\begin{aligned}
\mathbf{D}_{N\times}(m,u,w) &= \tfrac{\mathbf{D}_{\times}(m,u,w)-\mathbf{D}_\times^{min}}{\mathbf{D}_\times^{max}-\mathbf{D}_\times^{min}},\ \ \mathbf{E}_{N\times}(m,u,w) = \tfrac{\mathbf{E}_{\times}(m,u,w)-\mathbf{E}_\times^{min}}{\mathbf{E}_\times^{max}-\mathbf{E}_\times^{min}}\\
\mathbf{Q}_{N\times}(m,u,w) &= \tfrac{\mathbf{Q}_{\times}(m,u,w)-\mathbf{Q}_\times^{min}}{\mathbf{Q}_\times^{max}-\mathbf{Q}_\times^{min}},\ \
\mathbf{EQ}_{N\times}(m,u,w) = \tfrac{\mathbf{EQ}_{\times}(m,u,w)-\mathbf{EQ}_\times^{min}}{\mathbf{EQ}_\times^{max}-\mathbf{EQ}_\times^{min}},\\
\end{aligned}
\label{eq:normalized1}
\end{equation}
where $\times$ is either $A$ or $S$ and
\begin{equation}
\small
\begin{aligned}
\mathbf{D}_A^{\mathrm{m^{ax}_{in}}} &= \minmax\limits_{m,u,v,j}\left\{\tfrac{\widetilde{D}(m,u,w,j)}{N_{j}(m,u,w)}\right\},\ \ \mathbf{D}_S^{\mathrm{m^{ax}_{in}}} = \minmax\limits_{m,u,v,j}\left\{\widetilde{D}(m,u,w,j)\right\},\\
\mathbf{E}_A^{\mathrm{m^{ax}_{in}}} &= \minmax\limits_{m,u,v,j}\left\{\tfrac{\widetilde{E}(m,u,w,j)}{N_{j}(m,u,w)}\right\},\ \ \mathbf{E}_S^{\mathrm{m^{ax}_{in}}} = \minmax\limits_{m,u,v,j}\left\{\widetilde{E}(m,u,w,j)\right\},\\
\mathbf{Q}_A^{\mathrm{m^{ax}_{in}}} &= \minmax\limits_{m,u,v,j}\left\{\tfrac{\widetilde{Q}(m,u,w,j)}{N_{j}(m,u,w)}\right\},\ \ \mathbf{Q}_S^{\mathrm{m^{ax}_{in}}} = \minmax\limits_{m,u,v,j}\left\{\widetilde{Q}(m,u,w,j)\right\},\\
\mathbf{EQ}_A^{\mathrm{m^{ax}_{in}}} &= \minmax\limits_{m,u,v,j}\left\{\tfrac{\widetilde{EQ}(m,u,w,j)}{N_{j}(m,u,w)}\right\},\ \ \mathbf{EQ}_S^{\mathrm{m^{ax}_{in}}} = \minmax\limits_{m,u,v,j}\left\{\widetilde{EQ}(m,u,w,j)\right\},\\
\end{aligned}
\label{eq:normalized2}
\end{equation}
where $\mathrm{m^{ax}_{in}}$ is either $\min$ or $\max$.

\subsection{Moving strategies}

\label{sec:b_algorithms}

Let us assume a map with a desired ratio $S_1/S_2/S_3$ is ready and proceed to the algorithms expressing the dispersal strategies used in the computational experiments. The first strategy assumes no environment determination i.e. each animat moves freely without any constraints on where to go or when to stop.
\vspace{10pt}

\noindent{\bf Algorithm 2}: Dispersal without environment  determination ($u$=``Random'')
\begin{enumerate}
\item Let $(\mathcal{M},\mathcal{E})$ be a $L\times L$ map generated by Algorithm 1, where $\mathcal{M}$ shows the type of cells' strata and $\mathcal{E}$ shows the initial energy (attractiveness) of cells;
\item Let $A=\{a_1,\ldots,a_n\}$ be the set of animats; for each animat $a_k$ set its initial energy $E_k:=E^0$ and randomly choose
 a position $(x_k,y_k)$ from the positions $(x,y)$ with an appropriate starting quality: $Q_{min}\le M(x,y)\le Q_{max}$;  
\item Let $R_k$ be the array of the last $r$ cells visited by the animat $a_k$; initially let each $R_k$ be equal to $((x_k,y_k),(x_k,y_k),\ldots,(x_k,y_k))\in(\overline{1,L})^r$;
\item Repeat the following steps for $i\in\overline{1,T}$: 
\begin{enumerate}
\item[] For each animat $a_k\in A$:
\begin{enumerate}[(a)]
\item If $a_k$ is not engaged in a Levy walk:
\begin{enumerate}[i.]
    \item Choose a length $l_k$ of the next section of Levy walk randomly from $\overline{1,[L/2]}$ so that $$p\{l_k=x\}=\nicefrac{(1/x^2)}{\, \sum\limits_{j=1}^{[L/2]}1/j^2};$$
    \item If the previous direction vector of $a_k$ is not defined or if it is equal to zero, choose the direction vector of the next Levy walk section of $a_k$ arbitrary; otherwise choose it uniformly among all the vectors which dot product with the previous direction vector of $a_k$ is non-negative (the forward semiplane); 
\end{enumerate}
\item Choose the neighbor cell $(x_k',y_k')$ (among 9 possible in square grids) which center lies closest to the current movement direction; 
\item If $(x_k',y_k')$ lies outside of the map or $(x_k',y_k')$ is in $R_k$:
\begin{enumerate}[i.]
\item Interrupt the current section of Levy flight;
\item Reset the current direction vector to the zero vector (so that its forward semiplane will not affect the next direction);
\item Spend energy for the homeostasis: $E_k:=E_k - E^m/2$;
\end{enumerate}
\item Otherwise:
\begin{enumerate}[i.]
\item Change the position of $a_k$: $(x_k,y_k):=(x_k',y_k')$; 
\item Spend energy for the move: $E_k:=E_k - E^m$;
\end{enumerate}
\item Add the position $(x_k,y_k)$ at the first place in $R_k$, shift all the elements of $R_k$ such that the last one is pushed out of the memory;
\item Feed at the current cell considering the ``weather'' at the $i$-th step with the winter harshness $w$ (see \eqref{eq:weather_sine}): $$E_k:=E_k + \tfrac{1}{2}\ \mathcal{E}(x_k,y_k)\big((1-\tfrac{1}{w})\mathrm{cos}(\tfrac{2\pi i}{T})+(1+\tfrac{1}{w})\big);$$
\item If $E_k\le 0$, die, i.e. remove $a_k$ from $A$; 
\item If $|A|=\varnothing$, stop the algorithm.
\end{enumerate}
\end{enumerate}
\end{enumerate}

In all the experiments we take $L=100$, $n=100$, $T=1000$. The latter one is mostly a technical value which on one hand provides enough steps for 100 animats to distribute over a $100\times 100$ map and on the other -- allows to perform many experiment cycles in different conditions by keeping the computational time of a single experiment cycle relatively low. There are no restrictions on the starting cell quality: $Q_{min}=1, Q_{max}=3$. The energy $E^m$ spent by each animat for each move is taken equal to a half of the average energy of a ``medium'' cell: $E^m=25$. Recall that initial energy $E^0$ of each animat guarantees 10 moves without energy input: $E^0=10E^m$. Winter harshness $w$ takes on values in $\{2,4,6,8,10\}$. 

\vspace{10pt}

The next algorithm describes the ``Smart'' strategy of  movement (also known as {\it greedy} in optimization theory). The body of algorithm is mostly the same with Algorithm 2, so we only point out the differing item.
\vspace{10pt}

\noindent{\bf Algorithm 3}: Dispersal with full environment  determination ($u$=``Smart'')
\begin{enumerate}
\item[...] ... (see Algorithm 2) ... ...
\item[4(c).] If $(x_k',y_k')$ lies outside of the map or $(x_k',y_k')$ is in $R_k$ or $\mathcal{M}(x_k',y_k')<\mathcal{M}(x_k,y_k)$:
\item[...] ... (see Algorithm 2) ... ...
\end{enumerate}

\vspace{10pt}

It is {\it stop criterion} that distinguish the last -- ``Dreamer'' -- strategy: at each step ``Dreamer'' compares the attractiveness of its dream weighted by the chances to achieve it with the current habitat attractiveness. The subjective estimation of these chances decreases with every move where the animat does not find its dream-cell. Before proceeding to the algorithm let us go into the mathematical details of this decrease.

Let us imagine a series of experiments with the probability of success $p$ in each experiment. It is both intuitive and easy to prove that on average one needs to take $1/p$ steps until the first success: indeed, the expected value of the first successful step is
$$\mathbb{E}\triangleq 1\cdot p + 2(1-p)p + 3(1-p)^2p+\ldots;$$
turning the right part into a geometric series with the common ratio $1-p<1$:
$$(\mathbb{E}-p) - (1-p)\mathbb{E} = (1-p)p + (1-p)^2p + (1-p)^3p +\ldots = p\frac{1-p}{1-(1-p)}=1-p,$$
from where $\mathbb{E}=1/p$.  This evidence in the form $p=1/\mathbb{E}$ will help us to estimate the ``subjective chances'' of an animat to get to the dream. By construction the dream can not be achieved at the initial state. If an animat achieves its goal at the second step, then its estimation of the success probability will be $p_2=1/2$ (one success out of 2 cells: the first cell (initial) and the second (successful), so $\mathbb{E}=2$). If the goal is achieved for the first time at the third step, then a) the animat did not achieve it at the second step with probability $1-1/2$ (since, as we already mentioned, the subjective estimation of the success at the second step would have been 1/2) and b) the subjective estimation of success is one chance out of three moves $1/3$ ($\mathbb{E}=3$): the resulting subjective chances to get to the dream at the third step is $p_3=(1-1/2)/3$. Similarly, reaching the dream for the first time at the fourth step means that a) the dream was not reached at the second step (where the subjective probability of success would have been 1/2), b) was not reached at the third step (where the subjective success probability would have been 1/3) and c) was reached at the fourth ($\mathbb{E}=4$) with subjective probability 1/4, so the resulting subjective chance to achieve the dream at the fourth step is $p_4=(1-1/2)(1-1/3)/4$. Reasoning by analogy, the subjective probability to get to the dream at the $n$-th step for $n\ge3$ is 
$$p_n=\frac{1}{n}\prod_{i=3}^{n}\left(1-\frac{1}{i-1}\right).$$ 
The summary subjective chances of the animat to get to the dream in no more than $K\ge 3$ steps is 
\begin{equation}
\label{eq:part_sum}
S_K\triangleq\frac{1}{2}+\sum_{n=3}^K\left\{\frac{1}{n}\prod_{i=3}^{n}\left(1-\frac{1}{i-1}\right)\right\}=\sum_{n=1}^K\frac{1}{n(n+1)}
\end{equation}
(recall that $S_1\triangleq 0, S_2\triangleq 1/2$).
The introduced subjective chances function indeed resembles probability since it is known that $S_\infty\triangleq\lim_{K\rightarrow\infty}S_K=1$, which means that at the start an animat is fully determined to reach its dream.
Considering this and using the previous notations, the subjective chances to get to the dream after $K$ unsuccessful steps is $S_\infty - S_K = 1-S_K$. 
\vspace{10pt}

\noindent{\bf Algorithm 4}: Dispersal with partial environment  determination ($u$=``Dreamer'')
\begin{enumerate}
\item[...] ... (see Algorithm 2) ... ...
\item[$3'$.] Let $\mathcal{D}$ be the attractiveness of a dream cell;
\item[...] ... (see Algorithm 2) ... ...
\item[4(c).] If $(x_k',y_k')$ lies outside of the map, or $(x_k',y_k')$ is in $R_k$, or $$(S_\infty-S_K)\mathcal{D}<\tfrac{1}{2}\ \mathcal{E}(x_k,y_k)\big((1-\tfrac{1}{w})\mathrm{cos}(\tfrac{2\pi i}{T})+(1+\tfrac{1}{w})\big):$$
\item[...] ... (see Algorithm 2) ... ...
\end{enumerate}
\vspace{10pt}

The ``Dreamer'' stop condition is artificially included here into step 4(c) (at the expense of optimality) for the sake of clarity and uniformity with Algorithms 2 and 3. 

In our experiments we used  $\mathcal{D}=10000$. This value has a simple technical explanation: 1) the duration of each experiment cycle $T=1000$; 2) in the worst possible conditions -- ``bad'' map and the strongest winter $w=10$ -- the average initial cell energy is $0.67\cdot 10 + 0.3\cdot 50 + 0.03\cdot 90 = 24.4$ and the winter $w=10$ drops this value as low as $24.4/10=2.44$ at the peak with the average $(24.4+2.44)=13.42$; 3) with this average attractiveness of the cells, an animat with the dream $\mathcal{D}=10000$ will make on average 745 steps until its belief in dream bates enough for the animat to be satisfied with the average cell; 4) in the best possible conditions -- ``good'' map and $w=2$ -- average initial cell energy is $0.03\cdot 10 + 0.3\cdot 50 + 0.67\cdot 90 = 75.6$ and the average lowest is 75.6/2=37.8, which gives the total average cell energy of 56.7 and prompts an animat to make 175 steps on average until it rests. The provided speculations show that in all cases the ``Dreamer'' animats stop somewhere within the experiment cycle (from around the beginning to around the end).

\newpage
\section{Strategies graphs}
\label{sec:pictures}
\begingroup

\begingroup
\setlength{\tabcolsep}{15pt} 

\begin{figure}[H]
    \centering
\begin{tabular}{cccc}
\qquad & {\footnotesize``bad'' map} & {\footnotesize``medium'' map} & {\footnotesize``good'' map} \\
\rotatebox[origin=c]{90}{\footnotesize \quad\qquad``Random''\qquad\quad} &
\scalebox{0.7}{\raisebox{-.5\height}{\footnotesize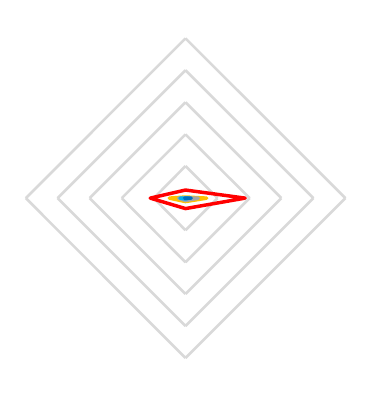}}
     &
\scalebox{0.7}{\raisebox{-.5\height}{\footnotesize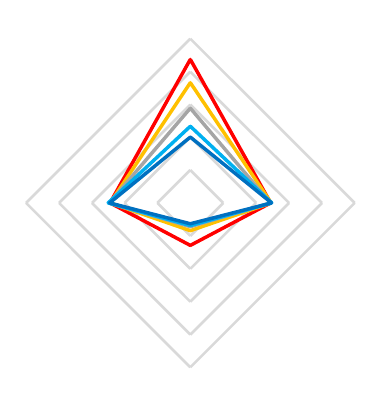}}
     &
\scalebox{0.7}{\raisebox{-.5\height}{\footnotesize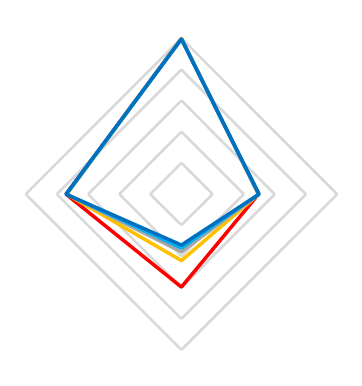}}
\\
\rotatebox[origin=c]{90}{\footnotesize \quad\qquad``Smart''\qquad\quad} &
\scalebox{0.7}{\raisebox{-.5\height}{\footnotesize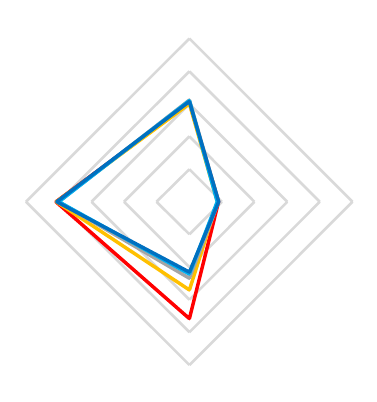}}
     &
\scalebox{0.7}{\raisebox{-.5\height}{\footnotesize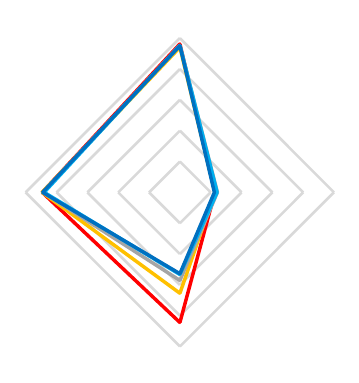}}
     &
\scalebox{0.7}{\raisebox{-.5\height}{\footnotesize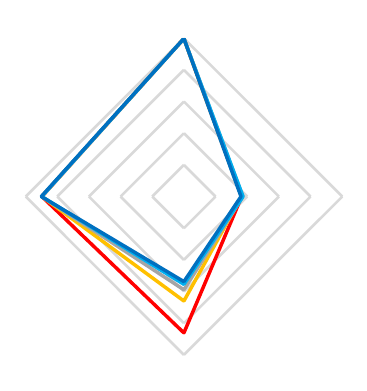}}
\\
\rotatebox[origin=c]{90}{\footnotesize \quad\qquad``Dreamer''\quad\qquad} &
\scalebox{0.7}{\raisebox{-.5\height}{\footnotesize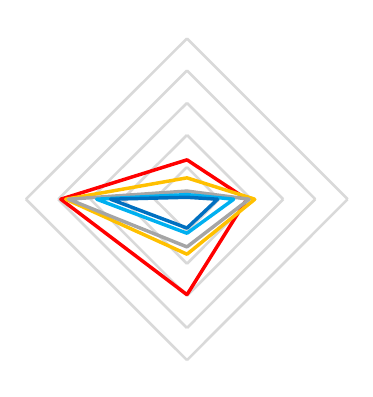}}
     &
\scalebox{0.7}{\raisebox{-.5\height}{\footnotesize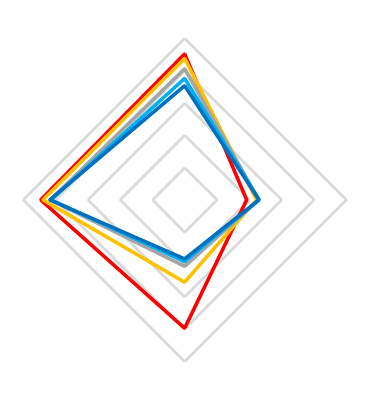}}
     &
\scalebox{0.7}{\raisebox{-.5\height}{\footnotesize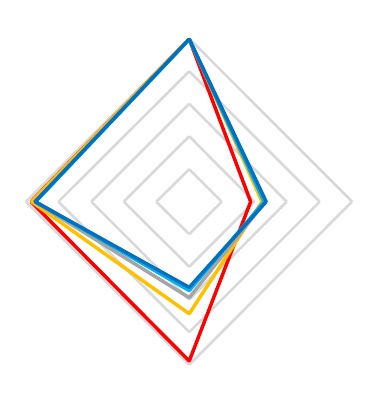}}
\\
\rotatebox[origin=c]{90}{\footnotesize \qquad All in one\quad\qquad} &
\scalebox{0.7}{\raisebox{-.5\height}{\footnotesize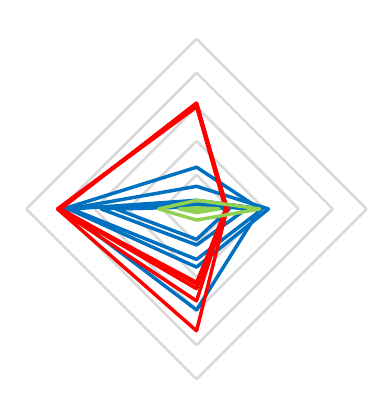}}
     &
\scalebox{0.7}{\raisebox{-.5\height}{\footnotesize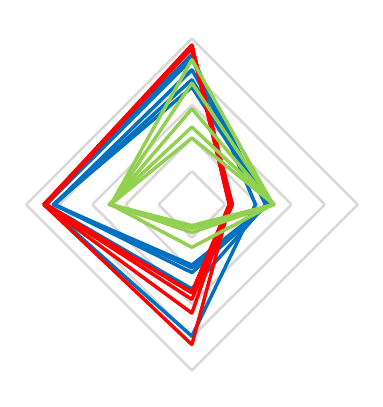}}
     &
\scalebox{0.7}{\raisebox{-.5\height}{\footnotesize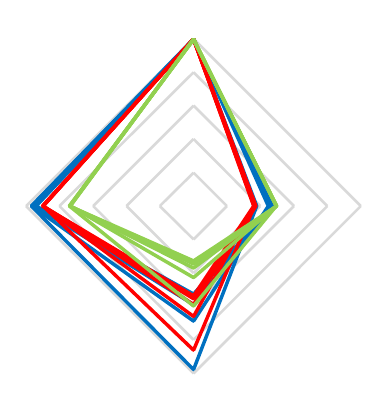}}
\end{tabular}
    \vspace{5mm}
    \caption{Portrays of relative performance of the strategies along different normalized average metrics \eqref{eq:normalized1}\eqref{eq:normalized2} (i.e. performance of the average survivors in the populations). In the first three rows, color shows the harshness of winter ($w$) in the rainbow order: from red ($w=2$) to dark blue ($w=10$). In the ``All in one'' row, color is used to mark off the strategies: green for ``Random'', red for ``Smart'' and blue for ``Dreamer''. All the axes run from 0 to 1 }
    \label{fig:radar_aver}
\end{figure}

\endgroup

\newpage

\begingroup
\setlength{\tabcolsep}{15pt} 

\begin{figure}[H]
    \centering
\begin{tabular}{cccc}
\qquad & {\footnotesize``bad'' map} & {\footnotesize``medium'' map} & {\footnotesize``good'' map} \\
\rotatebox[origin=c]{90}{\footnotesize \quad\qquad``Random''\qquad\quad} &
\scalebox{0.7}{\raisebox{-.5\height}{\footnotesize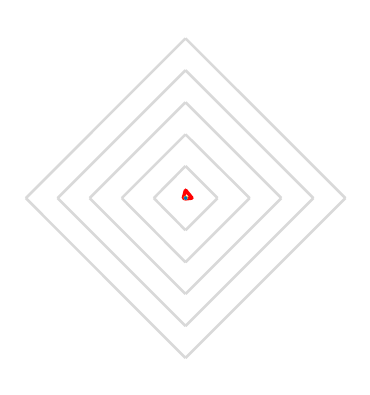}}
     &
\scalebox{0.7}{\raisebox{-.5\height}{\footnotesize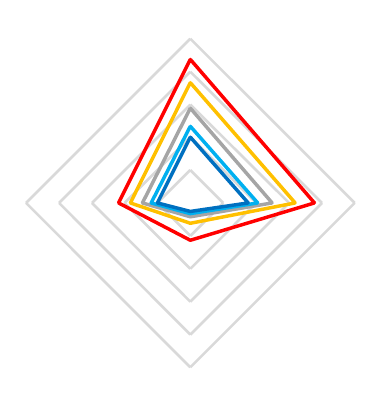}}
     &
\scalebox{0.7}{\raisebox{-.5\height}{\footnotesize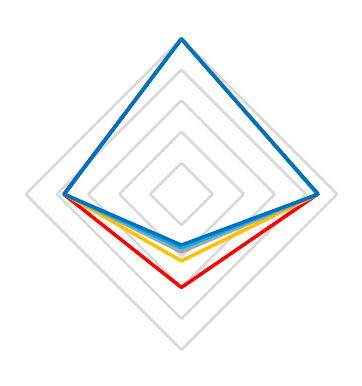}}
\\
\rotatebox[origin=c]{90}{\footnotesize \quad\qquad``Smart''\qquad\quad} &
\scalebox{0.7}{\raisebox{-.5\height}{\footnotesize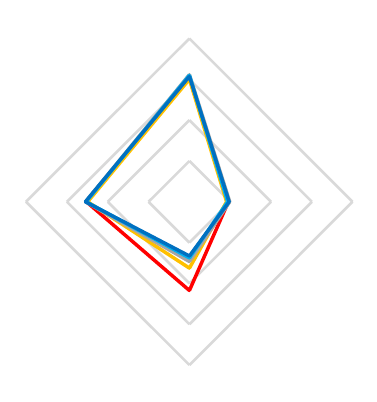}}
     &
\scalebox{0.7}{\raisebox{-.5\height}{\footnotesize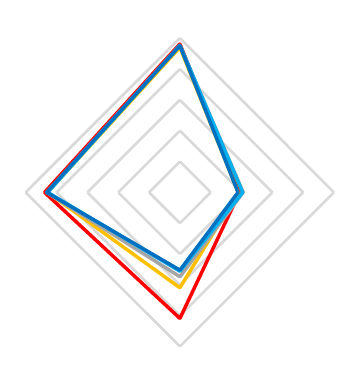}}
     &
\scalebox{0.7}{\raisebox{-.5\height}{\footnotesize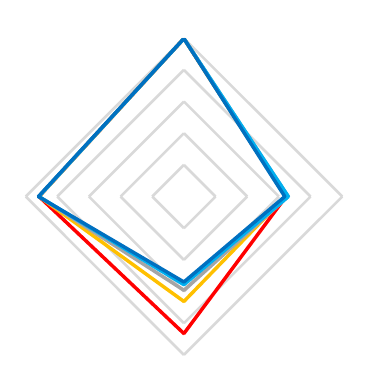}}
\\
\rotatebox[origin=c]{90}{\footnotesize \quad\qquad``Dreamer''\quad\qquad} &
\scalebox{0.7}{\raisebox{-.5\height}{\footnotesize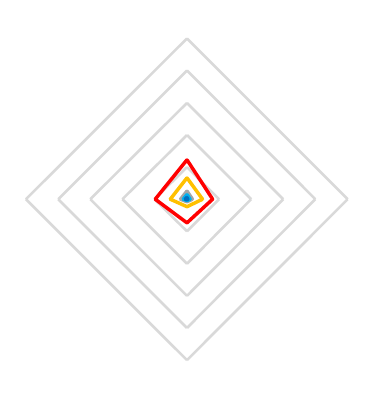}}
     &
\scalebox{0.7}{\raisebox{-.5\height}{\footnotesize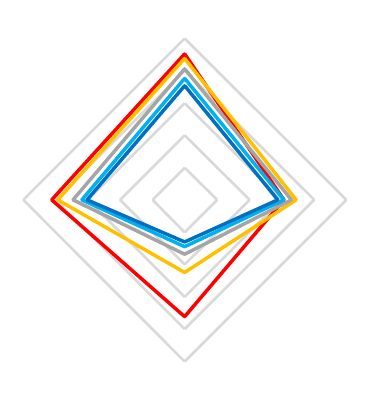}}
     &
\scalebox{0.7}{\raisebox{-.5\height}{\footnotesize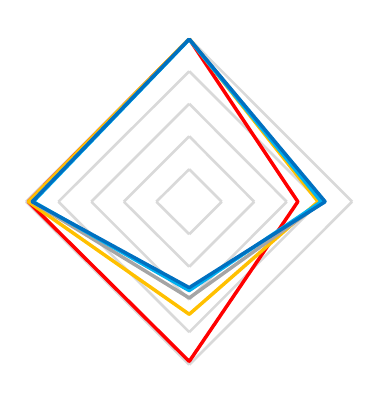}}
\\
\rotatebox[origin=c]{90}{\footnotesize \qquad All in one\quad\qquad} &
\scalebox{0.7}{\raisebox{-.5\height}{\footnotesize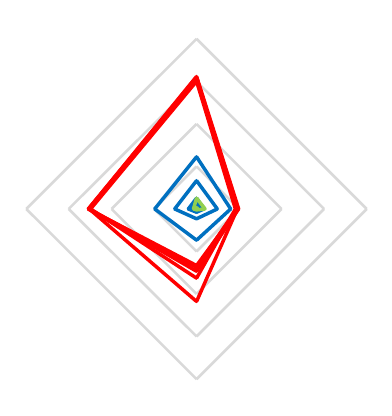}}
     &
\scalebox{0.7}{\raisebox{-.5\height}{\footnotesize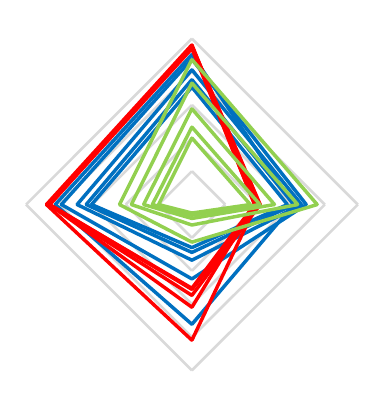}}
     &
\scalebox{0.7}{\raisebox{-.5\height}{\footnotesize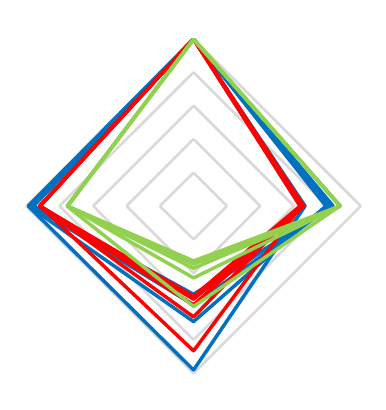}}
\end{tabular}
    \caption{Portrays of relative performance of the strategies along different normalized summary metrics \eqref{eq:normalized1}\eqref{eq:normalized2} (i.e. performance of the populations as wholes). In the first three rows, color shows the harshness of winter ($w$) in the rainbow order: from red ($w=2$) to dark blue ($w=10$). In the ``All in one'' row, color is used to mark off the strategies: green for ``Random'', red for ``Smart'' and blue for ``Dreamer''. All the axes run from 0 to 1 }
    \label{fig:radar_sum}
\end{figure}

\endgroup

\newpage

\bibliographystyle{model1-num-names}
\bibliography{bibl}

\begin{thebibliography}{36}
\expandafter\ifx\csname natexlab\endcsname\relax\def\natexlab#1{#1}\fi
\providecommand{\bibinfo}[2]{#2}
\ifx\xfnm\relax \def\xfnm[#1]{\unskip,\space#1}\fi
\bibitem[{Nathan et~al.(2008)Nathan, Getz, Revilla, Holyoak, Kadmon, Saltz, and
  Smouse}]{Nathan2008movement}
\bibinfo{author}{R.~Nathan}, \bibinfo{author}{W.~M. Getz},
  \bibinfo{author}{E.~Revilla}, \bibinfo{author}{M.~Holyoak},
  \bibinfo{author}{R.~Kadmon}, \bibinfo{author}{D.~Saltz},
  \bibinfo{author}{P.~E. Smouse},
\newblock \bibinfo{title}{A movement ecology paradigm for unifying organismal
  movement research},
\newblock \bibinfo{journal}{Proceedings of the {N}ational {A}cademy of
  {S}ciences of the {U}nited {S}tates of {A}merica} \bibinfo{volume}{105}
  (\bibinfo{year}{2008}) \bibinfo{pages}{19052--19059}.
\bibitem[{Earl et~al.(2016)Earl, Fuhlendorf, Haukos, Tanner, Elmore, and
  Carleton}]{Earl2016Characteristics}
\bibinfo{author}{J.~E. Earl}, \bibinfo{author}{S.~D. Fuhlendorf},
  \bibinfo{author}{D.~Haukos}, \bibinfo{author}{A.~M. Tanner},
  \bibinfo{author}{D.~Elmore}, \bibinfo{author}{S.~A. Carleton},
\newblock \bibinfo{title}{Characteristics of lesser prairie-chicken
  (\textit{{T}}{\it ympanuchus pallidicinctus}) long-distance movements across
  their distribution},
\newblock \bibinfo{journal}{Ecosphere} \bibinfo{volume}{7}
  (\bibinfo{year}{2016}) \bibinfo{pages}{1--13}.
\bibitem[{Fuglei and Tarroux(2019)}]{Fuglei2019Arctic}
\bibinfo{author}{E.~Fuglei}, \bibinfo{author}{A.~Tarroux},
\newblock \bibinfo{title}{Arctic fox dispersal from {S}valbard to {C}anada: One
  female’s long run across sea ice},
\newblock \bibinfo{journal}{Polar Research} \bibinfo{volume}{38}
  (\bibinfo{year}{2019}).
\bibitem[{Hindell and McMahon(2000)}]{Hindell2000Long}
\bibinfo{author}{M.~A. Hindell}, \bibinfo{author}{C.~R. McMahon},
\newblock \bibinfo{title}{Long distance movement of a southern elephant seal
  (\textit{{M}}{\it irounga leonina}) from {M}acquarie {I}sland to {P}eter 1
  {$\varnothing$y}},
\newblock \bibinfo{journal}{Marine Mammal Science} \bibinfo{volume}{16}
  (\bibinfo{year}{2000}) \bibinfo{pages}{504--507}.
\bibitem[{Weller et~al.(2016)Weller, Castle, Liechti, Hein, Schirmacher, and
  Cryan}]{Weller2016First}
\bibinfo{author}{T.~J. Weller}, \bibinfo{author}{K.~T. Castle},
  \bibinfo{author}{F.~Liechti}, \bibinfo{author}{C.~D. Hein},
  \bibinfo{author}{M.~R. Schirmacher}, \bibinfo{author}{P.~M. Cryan},
\newblock \bibinfo{title}{First direct evidence of long-distance seasonal
  movements and hibernation in a migratory bat},
\newblock \bibinfo{journal}{Scientific Reports} \bibinfo{volume}{6}
  (\bibinfo{year}{2016}) \bibinfo{pages}{1--7}.
\bibitem[{Truve et~al.(2004)Truve, Lemel, and Soderberg}]{Truve2004Dispersal}
\bibinfo{author}{J.~Truve}, \bibinfo{author}{J.~Lemel},
  \bibinfo{author}{B.~Soderberg},
\newblock \bibinfo{title}{Dispersal in relation to population density in wild
  boar (\textit{{S}}{\it us scrofa})},
\newblock \bibinfo{journal}{Galimys} \bibinfo{volume}{16}
  (\bibinfo{year}{2004}) \bibinfo{pages}{75--82}.
\bibitem[{Nathan et~al.(2003)Nathan, Perry, Cronin, Strand, Cain, and
  Methods}]{Nathan2003Methods}
\bibinfo{author}{R.~Nathan}, \bibinfo{author}{G.~Perry}, \bibinfo{author}{J.~T.
  Cronin}, \bibinfo{author}{A.~E. Strand}, \bibinfo{author}{M.~L. Cain},
  \bibinfo{author}{M.~L. Methods},
\newblock \bibinfo{title}{Methods for estimating long-distance dispersal},
\newblock \bibinfo{journal}{Oikos} \bibinfo{volume}{103} (\bibinfo{year}{2003})
  \bibinfo{pages}{261--273}.
\bibitem[{Hawkes(2009)}]{Hawkes2009Linking}
\bibinfo{author}{C.~Hawkes},
\newblock \bibinfo{title}{Linking movement behaviour, dispersal and population
  processes: Is individual variation a key?},
\newblock \bibinfo{journal}{Journal of Animal Ecology} \bibinfo{volume}{78}
  (\bibinfo{year}{2009}) \bibinfo{pages}{894--906}.
\bibitem[{Lewis et~al.(2015)Lewis, Petrovskii, and
  Potts}]{Lewis2015Mathematics}
\bibinfo{author}{M.~A. Lewis}, \bibinfo{author}{S.~V. Petrovskii},
  \bibinfo{author}{J.~R. Potts}, \bibinfo{title}{The Mathematics Behind
  Biological Invasions, Interdisciplinary Applied Mathematics},
  volume~\bibinfo{volume}{44}, \bibinfo{year}{2015}.
\bibitem[{Barton et~al.(2009)Barton, Phillips, Morales, and
  Travis}]{Barton2009evolution}
\bibinfo{author}{K.~A. Barton}, \bibinfo{author}{B.~L. Phillips},
  \bibinfo{author}{J.~M. Morales}, \bibinfo{author}{J.~M. Travis},
\newblock \bibinfo{title}{The evolution of an ``intelligent'' dispersal
  strategy: Biased, correlated random walks in patchy landscapes},
\newblock \bibinfo{journal}{Oikos} \bibinfo{volume}{118} (\bibinfo{year}{2009})
  \bibinfo{pages}{309--319}.
\bibitem[{Hiebeler(2004)}]{Hiebeler2004Competition}
\bibinfo{author}{D.~Hiebeler},
\newblock \bibinfo{title}{Competition between near and far dispersers in
  spatially structured habitats},
\newblock \bibinfo{journal}{Theoretical Population Biology}
  \bibinfo{volume}{66} (\bibinfo{year}{2004}) \bibinfo{pages}{205--218}.
\bibitem[{Davis and Stamps(2004)}]{Davis2004effect}
\bibinfo{author}{J.~M. Davis}, \bibinfo{author}{J.~A. Stamps},
\newblock \bibinfo{title}{The effect of natal experience on habitat
  preferences},
\newblock \bibinfo{journal}{Trends in Ecology and Evolution}
  \bibinfo{volume}{19} (\bibinfo{year}{2004}) \bibinfo{pages}{411--416}.
\bibitem[{Stamps et~al.(2005)Stamps, Krishnan, and Reid}]{Stamps2005Search}
\bibinfo{author}{J.~A. Stamps}, \bibinfo{author}{V.~V. Krishnan},
  \bibinfo{author}{M.~L. Reid},
\newblock \bibinfo{title}{Search costs and habitat selection by dispersers},
\newblock \bibinfo{journal}{Ecology} \bibinfo{volume}{86}
  (\bibinfo{year}{2005}) \bibinfo{pages}{510--518}.
\bibitem[{Stamps and Swaisgood(2007)}]{Stamps2007Someplace}
\bibinfo{author}{J.~A. Stamps}, \bibinfo{author}{R.~R. Swaisgood},
\newblock \bibinfo{title}{Someplace like home: Experience, habitat selection
  and conservation biology},
\newblock \bibinfo{journal}{Applied Animal Behaviour Science}
  \bibinfo{volume}{102} (\bibinfo{year}{2007}) \bibinfo{pages}{392--409}.
\bibitem[{Nathan(2005)}]{Nathan2005Long-distance}
\bibinfo{author}{R.~Nathan},
\newblock \bibinfo{title}{Long-distance dispersal research: Building a network
  of yellow brick roads},
\newblock \bibinfo{journal}{Diversity and Distributions} \bibinfo{volume}{11}
  (\bibinfo{year}{2005}) \bibinfo{pages}{125--130}.
\bibitem[{Sutherland et~al.(2000)Sutherland, Harestad, Price, and
  Lertzman}]{Sutherland2000Scaling}
\bibinfo{author}{G.~D. Sutherland}, \bibinfo{author}{A.~S. Harestad},
  \bibinfo{author}{K.~Price}, \bibinfo{author}{K.~P. Lertzman},
\newblock \bibinfo{title}{Scaling of natal dispersal distances in terrestrial
  birds and mammals},
\newblock \bibinfo{journal}{Ecology and Society} \bibinfo{volume}{4}
  (\bibinfo{year}{2000}).
\bibitem[{Teitelbaum et~al.(2015)Teitelbaum, Fagan, Fleming, Dressler,
  Calabrese, Leimgruber, and Mueller}]{Teitelbaum2015How}
\bibinfo{author}{C.~S. Teitelbaum}, \bibinfo{author}{W.~F. Fagan},
  \bibinfo{author}{C.~H. Fleming}, \bibinfo{author}{G.~Dressler},
  \bibinfo{author}{J.~M. Calabrese}, \bibinfo{author}{P.~Leimgruber},
  \bibinfo{author}{T.~Mueller},
\newblock \bibinfo{title}{How far to go? {D}eterminants of migration distance
  in land mammals},
\newblock \bibinfo{journal}{Ecology Letters} \bibinfo{volume}{18}
  (\bibinfo{year}{2015}) \bibinfo{pages}{545--552}.
\bibitem[{Wilson et~al.(2009)Wilson, Dormontt, Prentis, Lowe, and
  Richardson}]{Wilson2009Something}
\bibinfo{author}{J.~R. Wilson}, \bibinfo{author}{E.~E. Dormontt},
  \bibinfo{author}{P.~J. Prentis}, \bibinfo{author}{A.~J. Lowe},
  \bibinfo{author}{D.~M. Richardson},
\newblock \bibinfo{title}{Something in the way you move: dispersal pathways
  affect invasion success},
\newblock \bibinfo{journal}{Trends in Ecology and Evolution}
  \bibinfo{volume}{24} (\bibinfo{year}{2009}) \bibinfo{pages}{136--144}.
\bibitem[{Wilson(1991)}]{Wilson1991animat}
\bibinfo{author}{S.~Wilson},
\newblock \bibinfo{title}{The animat path to {AI}},
\newblock \bibinfo{journal}{Machine Learning}  (\bibinfo{year}{1991})
  \bibinfo{pages}{15--21}.
\bibitem[{Bertolero et~al.(2007)Bertolero, Oro, and
  Besnard}]{Bertolero2007Assessing}
\bibinfo{author}{A.~Bertolero}, \bibinfo{author}{D.~Oro},
  \bibinfo{author}{A.~Besnard},
\newblock \bibinfo{title}{Assessing the efficacy of reintroduction programmes
  by modelling adult survival: The example of {H}ermann's tortoise},
\newblock \bibinfo{journal}{Animal Conservation} \bibinfo{volume}{10}
  (\bibinfo{year}{2007}) \bibinfo{pages}{360--368}.
\bibitem[{Bremner-Harrison et~al.(2004)Bremner-Harrison, Prodohl, and
  Elwood}]{Bremner-Harrison2004Behavioural}
\bibinfo{author}{S.~Bremner-Harrison}, \bibinfo{author}{P.~A. Prodohl},
  \bibinfo{author}{R.~W. Elwood},
\newblock \bibinfo{title}{Behavioural trait assessment as a release criterion:
  Boldness predicts early death in a reintroduction programme of captive-bred
  swift fox (\textit{{V}ulpes velox})},
\newblock \bibinfo{journal}{Animal Conservation} \bibinfo{volume}{7}
  (\bibinfo{year}{2004}) \bibinfo{pages}{313--320}.
\bibitem[{Lodge(2003)}]{Lodge2003Biological}
\bibinfo{author}{D.~M. Lodge},
\newblock \bibinfo{title}{Biological invasions: Lessons for ecology},
\newblock \bibinfo{journal}{Trends in ecology and evolution}
  \bibinfo{volume}{8} (\bibinfo{year}{2003}) \bibinfo{pages}{133--136}.
\bibitem[{Viswanathan et~al.(1999)Viswanathan, Buldyrev, Havlin, Da~Luz,
  Raposo, and Stanley}]{Viswanathan1999Optimizing}
\bibinfo{author}{G.~M. Viswanathan}, \bibinfo{author}{V.~S. Buldyrev},
  \bibinfo{author}{S.~Havlin}, \bibinfo{author}{M.~G. Da~Luz},
  \bibinfo{author}{E.~P. Raposo}, \bibinfo{author}{H.~E. Stanley},
\newblock \bibinfo{title}{Optimizing the success of random searches},
\newblock \bibinfo{journal}{Nature} \bibinfo{volume}{401}
  (\bibinfo{year}{1999}) \bibinfo{pages}{911--914}.
\bibitem[{Gendreau and Potvin(2010)}]{metaheuristics2010handbook}
\bibinfo{author}{M.~Gendreau}, \bibinfo{author}{J.~Potvin},
  \bibinfo{title}{Handbook of Metaheuristics}, International Series in
  Operations Research \& Management Science, \bibinfo{publisher}{Springer US},
  \bibinfo{year}{2010}.
\bibitem[{Ivanko(2020)}]{software}
\bibinfo{author}{E.~Ivanko}, \bibinfo{title}{Kab{AN}imat: a tool for animals
  dispersal modelling},
  \bibinfo{howpublished}{\url{https://github.com/imm-complexity-lab/KabANimat}},
  \bibinfo{year}{2020}.
\bibitem[{Lotov and Miettinen(2008)}]{pareto}
\bibinfo{author}{A.~Lotov}, \bibinfo{author}{K.~Miettinen},
\newblock \bibinfo{title}{Visualizing the {P}areto frontier},
\newblock in: \bibinfo{editor}{J.~Branke}, \bibinfo{editor}{K.~Deb},
  \bibinfo{editor}{K.~Miettinen}, \bibinfo{editor}{R.~Slowinski} (Eds.),
  \bibinfo{booktitle}{Multiobjective Optimization, {LNCS}}, volume
  \bibinfo{volume}{5252}, \bibinfo{publisher}{Springer},
  \bibinfo{address}{Berlin, Heidelberg}, \bibinfo{year}{2008}, pp.
  \bibinfo{pages}{213--243}.
\bibitem[{Bowler and Benton(2005)}]{Bowler2005Causes}
\bibinfo{author}{D.~E. Bowler}, \bibinfo{author}{T.~G. Benton},
\newblock \bibinfo{title}{Causes and consequences of animal dispersal
  strategies: Relating individual behaviour to spatial dynamics},
\newblock \bibinfo{journal}{Biological Reviews of the {C}ambridge
  {P}hilosophical {S}ociety} \bibinfo{volume}{80} (\bibinfo{year}{2005})
  \bibinfo{pages}{205--225}.
\bibitem[{Kubisch et~al.(2014)Kubisch, Holt, Poethke, and
  Fronhofer}]{Kubisch2014Where}
\bibinfo{author}{A.~Kubisch}, \bibinfo{author}{R.~D. Holt},
  \bibinfo{author}{H.~J. Poethke}, \bibinfo{author}{E.~A. Fronhofer},
\newblock \bibinfo{title}{Where am i and why? {S}ynthesizing range biology and
  the eco-evolutionary dynamics of dispersal},
\newblock \bibinfo{journal}{Oikos} \bibinfo{volume}{123} (\bibinfo{year}{2014})
  \bibinfo{pages}{5--22}.
\bibitem[{Ramanantoanina et~al.(2014)Ramanantoanina, Ouhinou, and
  Hui}]{Ramanantoanina2014Spatial}
\bibinfo{author}{A.~Ramanantoanina}, \bibinfo{author}{A.~Ouhinou},
  \bibinfo{author}{C.~Hui},
\newblock \bibinfo{title}{Spatial assortment of mixed propagules explains the
  acceleration of range expansion},
\newblock \bibinfo{journal}{PLoS ONE} \bibinfo{volume}{9}
  (\bibinfo{year}{2014}).
\bibitem[{Zollner and Lima(1999)}]{Zollner1999Search}
\bibinfo{author}{P.~A. Zollner}, \bibinfo{author}{S.~L. Lima},
\newblock \bibinfo{title}{Search strategies for landscape-level interpatch
  movements},
\newblock \bibinfo{journal}{Ecology} \bibinfo{volume}{80}
  (\bibinfo{year}{1999}) \bibinfo{pages}{1019--1030}.
\bibitem[{Kokko and Sutherland(2001)}]{Kokko2001Ecological}
\bibinfo{author}{H.~Kokko}, \bibinfo{author}{W.~J. Sutherland},
\newblock \bibinfo{title}{Ecological traps in changing environments: Ecological
  and evolutionary consequences of a behaviourally mediated {A}llee effect},
\newblock \bibinfo{journal}{Evolutionary Ecology Research} \bibinfo{volume}{3}
  (\bibinfo{year}{2001}) \bibinfo{pages}{537--551}.
\bibitem[{Conradt et~al.(2003)Conradt, Zollner, Roper, Frank, and
  Thomas}]{Conradt2003Foray}
\bibinfo{author}{L.~Conradt}, \bibinfo{author}{P.~A. Zollner},
  \bibinfo{author}{T.~J. Roper}, \bibinfo{author}{K.~Frank},
  \bibinfo{author}{C.~D. Thomas},
\newblock \bibinfo{title}{Foray search: An effective systematic dispersal
  strategy in fragmented landscapes},
\newblock \bibinfo{journal}{American Naturalist} \bibinfo{volume}{161}
  (\bibinfo{year}{2003}) \bibinfo{pages}{905--915}.
\bibitem[{Shigesada and Kawasaki(2002)}]{Shigesada2002Invasion}
\bibinfo{author}{N.~Shigesada}, \bibinfo{author}{K.~Kawasaki},
\newblock \bibinfo{title}{Invasion and the range expansion of species: effects
  of long-distance dispersal},
\newblock in: \bibinfo{editor}{J.~Bullock}, \bibinfo{editor}{K.~R. E.},
  \bibinfo{editor}{R.~S. Hails} (Eds.), \bibinfo{booktitle}{Dispersal Ecology:
  the 42 Symposium of the British Ecological Society},
  \bibinfo{publisher}{Blackwell Publishing}, \bibinfo{year}{2002}, pp.
  \bibinfo{pages}{350--373}.
\bibitem[{Jung and Larter(2017)}]{Jung2017Observations}
\bibinfo{author}{T.~S. Jung}, \bibinfo{author}{N.~C. Larter},
\newblock \bibinfo{title}{Observations of long-distance post-release dispersal
  by reintroduced bison (\textit{{B}ison bison})},
\newblock \bibinfo{journal}{The Canadian Field-Naturalist}
  \bibinfo{volume}{131} (\bibinfo{year}{2017}) \bibinfo{pages}{221--224}.
\bibitem[{Reale et~al.(2007)Reale, Reader, Sol, McDougall, and
  Dingemanse}]{Reale2007Integrating}
\bibinfo{author}{D.~Reale}, \bibinfo{author}{S.~M. Reader},
  \bibinfo{author}{D.~Sol}, \bibinfo{author}{P.~T. McDougall},
  \bibinfo{author}{N.~J. Dingemanse},
\newblock \bibinfo{title}{Integrating animal temperament within ecology and
  evolution},
\newblock \bibinfo{journal}{Biological Reviews} \bibinfo{volume}{82}
  (\bibinfo{year}{2007}) \bibinfo{pages}{291--318}.
\bibitem[{Malange et~al.(2016)Malange, Izar, and
  Japyassu}]{Malange2016Personality}
\bibinfo{author}{J.~Malange}, \bibinfo{author}{P.~Izar},
  \bibinfo{author}{H.~Japyassu},
\newblock \bibinfo{title}{Personality and behavioural syndrome in
  \textit{{N}ecromys lasiurus} ({R}odentia: {C}ricetidae): notes on dispersal
  and invasion processes},
\newblock \bibinfo{journal}{Acta Ethologica} \bibinfo{volume}{19}
  (\bibinfo{year}{2016}) \bibinfo{pages}{189--195}.

\end{thebibliography}







\end{document}